\documentclass[aps, prd, nofootinbib,
    amsmath, a4paper, 10pt, twocolumn, 
    showkeys, superscriptaddress,preprintnumbers]{revtex4-2}
\usepackage[english]{babel}
\usepackage[margin=2cm]{geometry}
\usepackage{listings}
\usepackage{graphicx}
\usepackage{amssymb, amsmath}
\usepackage{float}
\usepackage{hyperref}
\usepackage{booktabs}
\usepackage{adjustbox}
\usepackage{rotating}
\usepackage[dvipsnames]{xcolor}
\providecommand{\margbf}[2]{\ensuremath{#1\,\bigl(#2\bigr)}}
\providecommand{\notapp}{\ensuremath{\text{--}}}

\definecolor{dourado}{rgb}{0.56, 0.45, 0.0}

\begin{document}

\title{Stronger-to-weaker gravity in generalized Brans–Dicke cosmology and implications for the Hubble and $S_8$ tensions}

\author{Lucas P. C. Leão}
\affiliation{Departamento de F\'isica, Universidade Federal de Minas Gerais (UFMG), Belo Horizonte, MG, Brazil}
\author{Gláuber C. Dorsch}
\affiliation{Departamento de F\'isica, Universidade Federal de Minas Gerais (UFMG), Belo Horizonte, MG, Brazil}
\author{Sergio E. Jorás}
\affiliation{Departamento de F\'isica, Universidade Federal do Rio de Janeiro (UFRJ), Rio de Janeiro, RJ, Brazil}
\author{Gonzalo J. Olmo}
\affiliation{Instituto de Física Corpuscular (IFIC), CSIC-Universitat de València, València, Spain}
\affiliation{Departamento de F\'\i sica, Universidade Federal da Para\'\i ba, 58051-970 Jo\~ao Pessoa, PB, Brazil}

\begin{abstract}
We investigate the cosmological implications of a Brans--Dicke-type scalar-tensor theory with non-minimal coupling $A(\phi)=1+c_1\phi+c_2\phi^2$ and power-law potentials $V(\phi)\propto\phi^{2n}$ ($n=2,3,4$), in which the dynamics of the scalar field yields a time-varying effective gravitational strength $G_{\rm eff}$. The theory is restricted to ghost-free solutions that also satisfy the dominant energy condition, sub-Planckian field ranges, as well as solar-system and big-bang nucleosynthesis bounds, keeping the scalar sector in a well-controlled and motivated regime. When confronting the model with Planck CMB, DESI DR2 BAO and Pantheon+ supernova data with the SH0ES calibration, the preferred scalar field dynamics lowers the sound horizon and raises the Hubble constant to $H_0=71.09\pm0.66\,{\rm km\,s^{-1}\,Mpc^{-1}}$ for $n=3$, which also shows a moderate Bayesian preference over $\Lambda$CDM ($\ln B=4.55\pm0.44$) under the adopted priors. We show that, within this framework, the early stronger-gravity and late weaker-gravity phases compete for the same coefficients, forcing the latter to have amplitudes bounded by $1-G_{\rm eff}/G \lesssim 10^{-4}$. The data also prefer only brief weaker-gravity excursions after recombination for $n=2$ and $n=3$, lasting less than $0.1$ e-folds, but $n=4$ shows late-time transitions at $z\lesssim 3$ persisting for up to $\sim 1.9$ e-folds. In all cases the weaker-gravity regime is far too shallow to suppress the growth of structure: the $S_8$ parameter is in fact higher than in $\Lambda$CDM and the preferred cosmologies are of the early-modified-gravity type.
\end{abstract}
\maketitle

\section{Introduction}
Along the past few decades the $\Lambda$CDM model has been extremely successful in describing the overall features of the cosmic history. Nevertheless, recent percent-level precision data on the Cosmic Microwave Background (CMB)~\cite{Planck:2018vyg, AtacamaCosmologyTelescope:2025blo, SPT-3G:2025bzu}, confronted with late-time data on Baryon Acoustic Oscillations (BAO) + Large-Scale Structure (LSS)~\cite{DESI:2025zgx} and supernovae~\cite{Brout:2022vxf} have placed this simple model under severe tensions~\cite{Abdalla:2022yfr,Perivolaropoulos:2021jda,CosmoVerse:2025}. 
The most significant of all is the Hubble tension, where direct measurements of the Hubble constant $H_0$ at low redshift disagree with inferences from the CMB measurements at $\gtrsim 5\sigma$ level~\cite{Riess:2021jrx,DiValentino:2021izs,Freedman:2024eph}. This tension is often accompanied by discrepancies in measurements of the distribution of matter, encapsulated in the $S_8\equiv \sigma_8 \sqrt{\Omega_m/0.3}$ parameter and the growth index $\gamma$ (defined via the growth rate approximation $f \simeq \Omega_m(a)^\gamma$~\cite{Linder:2005in,Nguyen:2023fip}): from the early-time CMB measurements one infers a more clumped matter-distribution than obtained from late-time weak-lensing and galaxy surveys, the tension amounting to a $2$--$3\sigma$ discrepancy in $S_8$~\cite{KiDS:2020suj,DES:2021wwk}.

In attempts to alleviate both tensions, several amendments have been proposed on the standard model over the last decade (cf.~\cite{Abdalla:2022yfr,Poulin:2023lkg,CosmoVerse:2025} for reviews). One intriguing class of possibilities consists in postulating a modified gravity scenario where the effective gravitational strength ``constant'' $G_\text{N}$ varies in time. If gravity were stronger in the early Universe than it is today, namely if $G_\text{N}>G$ before the baryonic fluid decouples from radiation,
then the size of the sound horizon at this decoupling would be smaller than predicted by $\Lambda$CDM~\cite{Zumalac_rregui_2020,Ballesteros_2020,Braglia:2020iik,Ballardini:2020iws,Braglia:2020auw,Ye:2024ywg}. This sound-horizon, given by $r_{\rm drag} = \int_{z_d}^\infty c_s(z)/H(z)$ (with $c_s(z)$ being the speed-of-sound of the baryon-photon fluid, $H(z)$ the Hubble parameter, and $z_d$ the redshift of the drag epoch), is directly obtained from the angular size of the first peak in the CMB spectrum, $\theta_s$, as well as the distance $D_A=\int_{z_d}^0 dz/H(z)$ traveled by the CMB photon from the last-scattering-surface until today. Hence a smaller sound horizon, implied by a modified Hubble history before baryon-photon decoupling, may lead to a larger Hubble rate as inferred from the CMB, which is what we need to alleviate the Hubble tension. However, increasing the expansion rate before recombination is usually followed by a higher dark matter physical density, $\Omega_{\rm cdm}h^2$, to preserve the peak heights, which then leads to more matter clustering at late-times and worsens the $S_8$ tension~\cite{Hill:2020osr,Ivanov:2020ril,Jedamzik:2020zmd}. If one wants to solve the $S_8$ tension in this paradigm, one actually needs gravity to be weaker at late times~\cite{Marra:2021fvf,Farhang:2020sij,Khosravi:2021csn,Heisenberg:2022gqk}. 

The impact of such a time-varying gravitational coupling on the $H_0-S_8$ tensions has been investigated by imposing the stronger-to-weaker gravity phenomenologically~\cite{Marra:2021fvf,Farhang:2020sij,Benevento:2022cql,Kable:2023} or within effective-field-theory descriptions of modified gravity~\cite{Perenon:2015sla,Gannouji:2020ylf}. However, a time-evolving gravitational coupling can also appear naturally in a scalar-tensor framework of modified gravity~\cite{Zumalac_rregui_2020,Ballesteros_2020,Braglia:2020auw,Ye:2024ywg}. The goal of this work is to explore a covariant scalar-tensor theory realizing this desired behavior for $G_\text{N}$ while also obeying  stability (ghost-free) and the dominant energy conditions, as well as sub-Planckian field excursions. We argue that a Generalized Brans-Dicke formulation~\cite{Brans:1961sx} with a polynomial non-minimal coupling function $A(\phi)=1+c_1\phi+c_2\phi^2$ and a field potential $V(\phi)\propto \phi^{2n}$ has the desired behavior, and explore its parameter space under confrontation with the data from {Planck 2018}, DESI DR2 BAO, and Pantheon+ with SH0ES.

We find that the theoretical and local conditions alone already shape the phenomenology of the model, with the early stronger-gravity and late weaker-gravity phases competing for the same coupling coefficients $c_1$ and $c_2$. A percent-level early enhancement of the gravitational coupling limits the weaker-gravity regime to $1-G_{\rm eff}/G\lesssim10^{-4}$, while its epoch and duration can only be determined by cosmological data. However, Planck and DESI data alone do not distinguish the model from $\Lambda$CDM, although more specific constraints are found once the SH0ES-calibrated supernova data are included. In particular, the scalar field becomes dynamical near matter--radiation equality, effectively lowering the sound horizon and raising the Hubble constant to $H_0=71.09\pm0.66\,\mathrm{km\,s^{-1}\,Mpc^{-1}}$ for $n=3$, with a moderate Bayesian preference over $\Lambda$CDM. On the other hand, the data realize the weaker-gravity regime either as brief excursions after recombination ($n=2,3$) or as an abrupt late-time transition ($n=4$), but always at a negligible depth, so that $S_8$ is not reduced and the preferred cosmologies are of the early-modified-gravity type. An important conclusion is that a more elaborate construction is therefore needed in order to solve both the $H_0$ and the $S_8$ tensions simultaneously. Nevertheless, we believe the current work is an important step in determining the key ingredients to a full answer for both tensions\footnote{Besides, note that recent analyses with an improved calibration of weak-lensing data (see e.g.~\cite{Wright:2025xka}), or with a stronger suppression of power on non-linear scales, e.g. due to baryonic feedback~\cite{Amon:2022azi,Preston:2023uup}, have considerably reduced or even removed the tension on the growth of structure.}.

The paper is organized as follows: In Sec.~II, we introduce the theoretical framework, establish the field equations, and derive initial bounds from energy and stability conditions. Sec.~III discusses the effective gravitational constant, PPN parameters, and solar system or BBN constraints, and estimates the maximum depth of the weaker-gravity regime. Sec.~IV details the parameter space exploration using Sobol sequences and the statistical settings for our nested sampling runs. Our main cosmological results, parameter constraints, and the behavior of the scalar field, the weaker-gravity phase and the matter power spectrum are presented in Sec.~V. Finally, Sec.~VI is dedicated to our conclusions and final remarks.

\section{Theoretical Framework and Phenomenological Motivations}
\label{sec:theory}

We consider a subclass of Horndeski gravity defined by the action 
\begin{align}
    S = \int d^4x \sqrt{-g} \Bigg[\frac{A(\phi)}{2} R - \frac{1}{2} \partial_\mu \phi \partial^\mu \phi - V(\phi) \Bigg] + S_m, \label{eq:original_action}
\end{align}
with a non-minimal coupling function $A(\phi)$ and a power-law potential $V(\phi)$ given by
\begin{align}
    A(\phi) &= 1 + c_1\phi + c_2\phi^2 ,\label{eq:coupling}\\
    V(\phi) &= \frac{V_0\phi^{2n}}{1.45\times 10^{113}}m^4_{\rm Pl} \label{eq:potential},
\end{align}
where $c_1, c_2, \phi, V_0$ are all dimensionless quantities, and we work in units of the reduced Planck mass $m_\text{Pl} = 1/\sqrt{8\pi G} = 1$. The denominator of the potential is a conversion factor from $m^2_{\rm Pl}$ to Mpc$^{-2}$, which is necessary to agree with the unit convention of the \texttt{hi\_class} \cite{Zumalac_rregui_2017, Bellini_2020} Boltzmann solver used in this work\footnote{hi\_class makes the Lagrangian dimensionless by dividing it by $m_{\rm Pl}^2L^{-2}$, where $L=1\,\mathrm{Mpc}$. The dimensionless amplitude $V_0$ is the physical amplitude of the potential, $V_{\rm phys}$ (dimension of energy$^4$), divided by this same factor, $V_0=V_{\rm phys}L^2/m_{\rm Pl}^2$. Hence, in Planck units, $V_{\rm phys}/m_{\rm Pl}^4=V_0/(m_{\rm Pl}L)^2=V_0/(1.45\times10^{113})$.}.

The coupling function can be viewed as the joint action of a Brans-Dicke-like linear term, $c_1\phi R$, and a Higgs-like non-minimal quadratic coupling, $(1+c_2\phi^2)R$, both of which are broadly studied actions on their own (see, e.g., \cite{Damour:1992kf, Acquaviva_2005, Avilez_2014, Peracaula_2019} for the former and \cite{Braglia:2020iik, Ballardini:2020iws} for the latter)\footnote{Notice that a field shift $\psi = \phi +  c_1 / 2c_2$ can eliminate the linear term and cast the coupling function in the form $A(\psi) = 1 - (c_1^2/4c_2) + c_2\psi^2$. This form is a particular case of a generally coupled scalar-tensor theory \cite{PhysRevD.54.1460, 1997CQGra_14_1981H} and recovers the $\mathbb{Z}_2$ symmetry on the coupling function, but with the potential minimum also shifted to $\psi = c_1 / 2c_2$. Although equivalent, the formulation proposed in eqs.~\eqref{eq:coupling},~\eqref{eq:potential} is more intuitive, as it can be viewed as a power-series expansion around zero truncated at order 2, which is a reasonable approximation for a wide variety of more complex functions.}. For theories described by eq.~\eqref{eq:original_action}, the positivity of the effective Planck mass squared, $\tilde{m}_\text{Pl}^2(\phi) \equiv m_\text{Pl}^2A(\phi)$ and scalar kinetic terms requires \cite{Gannouji_2006,Braglia:2020iik}
\begin{equation}
A>0,
\qquad
A\left(2A+3A_{,\phi}^{2}\right)>0.
\end{equation}
Notice that the second condition is automatically satisfied whenever $A>0$ and the scalar field is real, since then $A_{,\phi}^{2}\geq0$. 

To further understand the dynamics produced by action~\eqref{eq:original_action}, we vary it with respect to $\phi$ and to the metric to obtain the Klein-Gordon and 1st Friedmann equations:
\begin{align}
  &\ddot{\phi} + 3H\dot{\phi} + V_{,\phi} - \tfrac{1}{2}A_{,\phi}R = 0, \label{eq:KG}\\
  &3H^2 = \rho_{\rm std} + \rho_\phi, \label{eq:friedmann}
\end{align}
where $\rho_{\rm std}$ is the energy density of the standard components (matter, radiation and the cosmological constant) and
\begin{equation}
    \rho_\phi \equiv \frac{1}{2}\dot{\phi}^2 + V(\phi) - 3H\left[ \dot{\phi}A_{,\phi} + H(A-1) \right]\label{eq:energy_density}
\end{equation}
is the energy density of the scalar field.

Eq.~\eqref{eq:KG} has the form of a damped oscillator: $3H\dot\phi$ is the Hubble friction, while $\tfrac12 A_{,\phi}R - V_{,\phi}$ is the restoring force. As long as $H^2$ is much larger than such force, as it happens on the very early Universe, friction dominates and the field remains frozen at its initial value. Then, as $H$ decreases monotonically, the potential eventually overcomes the friction and the field thaws, rolls down the potential and oscillates around its minimum, with the friction progressively damping the oscillations; the first panel of fig.~\ref{fig:bestfit_scalar_background} illustrates this behavior.

Treating the scalar sector as an effective perfect fluid, we can apply some reasonable requirements to find initial bounds on $c_1$ and $c_2$. First, at the background level, it is desirable --- though not mandatory, since the field is not an actual fluid and it does not dominate the universe expansion --- for the scalar energy-momentum content to satisfy the dominant energy condition (DEC), which determines that the energy-momentum flux measured by a timelike observer is causal \cite{Kontou_2020}. In practice, the DEC manifests as
\begin{align}
\mathrm{DEC}: \qquad&
\rho_{\phi}\geq\left|p_{\phi}\right|,
\end{align}
further demanding that $\rho_\phi \geq 0$ and $\rho_\phi \geq -p_\phi$. For a positive energy density, these combined effects are equivalent to
\begin{equation}
-1\leq w_{\phi}\equiv\frac{p_{\phi}}{\rho_{\phi}}\leq1 \qquad \text{for } \rho_\phi >0,
\end{equation}
which automatically enforces the weak and null energy conditions (WEC and NEC) as well. The WEC states that timelike observers measure a non-negative effective energy density, and the NEC excludes phantom-like backgrounds. In ordinary two-derivative phantom-scalar models, NEC violation is associated with ghosts and rapid vacuum decay, and more generally with instabilities or superluminal modes in broad classes of cosmological fluids and fields \cite{Carroll_2003,Buniy_2006,Dubovsky_2006}. It is worth noticing that controlled higher-derivative exceptions are known \cite{Creminelli_2006,Rubakov_2014}, so we adopt these inequalities as conservative background-level priors rather than universal stability theorems.

The non-negativity of the energy density is particularly relevant to motivate the inclusion of the quadratic term, $c_2\phi^2$, in the coupling function. The desire to study a stronger-to-weaker gravitational strength ultimately requires a transition from $A(\phi) < 1$ to $A(\phi) > 1$, therefore $c_1 \neq 0$ is strictly needed for a coupling truncated at second order. This, however, places no constraint on $c_2$. In principle, the quadratic term is not necessary for the transition, but if we were to restrict the model to a purely linear coupling by setting $c_2 = 0$, the necessary oscillations of the scalar field around zero, together with the DEC, force $c_1$ to be very small and suppress the new physics. To see this, we start from the energy density of the scalar field given by eq.~\eqref{eq:energy_density}, which in a model with $c_2=0$ takes the form
\begin{align}
    \rho_\phi&= \frac{1}{2}\dot{\phi}^2 + V_0\phi^{2n}- 3c_1H\left(\dot{\phi} + H\phi\right)
\end{align}
The $\rho_\phi \geq 0$ condition then forces the general constraint
\begin{equation}\label{eq:condition_general}
    \frac{1}{6H}\left(2V_0\phi^{2n}+\dot{\phi}^2\right) \geq  c_1(\dot{\phi} + H\phi).
\end{equation}

When $\phi$ crosses zero (otherwise the gravity transition does not occur), the potential vanishes ($V(0)=0$), and all the field's energy is kinetic. At this exact crossing point, the $\rho_\phi \geq 0$ condition is only respected if
\begin{equation}\label{eq:constraint_c1}
    |c_1| \leq \frac{|\dot{\phi}_v|}{6H}.
\end{equation}
where $\dot{\phi}_v$ is the field's velocity at the crossing. If the field unfreezes at early times (as is required for early modified gravity), the Hubble parameter $H$ is typically much greater than the field velocity, forcing $|c_1| \ll 1$. Consequently, if we assume $\phi_{\rm ini} \sim \mathcal{O}(1)$, the modifications to gravity are highly suppressed, and no interesting new physics can be expected.

These obstacles are the primary motivation for introducing the $c_2\phi^2$ term: it allows transitions to weaker-gravity without $\phi$ crossing zero and, since $c_2\phi^2$ and its derivative naturally vanish at a $\phi=0$, the modulus $|c_2|$ remains relatively unconstrained by the $\rho_\phi > 0$ condition. Nevertheless, eq.~\eqref{eq:energy_density} strongly limits the positivity of $c_2$. At very early times, the Hubble parameter dominates the other terms and $\rho_\phi \sim H^2[1-A(\phi_\text{ini})]$, therefore $A(\phi_\text{ini}) < 1$ is strictly required and a positive $c_2$ cannot be arbitrarily large, as it pushes $A(\phi_\text{ini})$ to higher values. The extent of this constraint will depend on the allowed values of $c_1$.

The lack of $\mathbb{Z}_2$ symmetry in this model requires a choice of sign for the initial conditions, and we choose to work with $\phi_{\rm ini} > 0$. Under the full functional form of eq.~\eqref{eq:coupling}, the requirement that $0 < A(\phi_{\rm ini}) < 1$ translates to the algebraic constraint:
\begin{equation}\label{eq:condition_strength}
    -1 < c_1\phi_{\rm ini} + c_2\phi_{\rm ini}^2 < 0.
\end{equation}
Unlike the dynamic bounds of eq.~\eqref{eq:condition_general}, which depend on the full evolution of $H(t)$ and $\dot{\phi}(t)$ and must be checked numerically, the constraint in eq.~\eqref{eq:condition_strength} can be implemented at the input level to immediately reject undesirable regions of the parameter space.

On another note, we also impose the separate field-range prior
\begin{equation}
\phi^{2}<A(\phi),
\end{equation}
which restricts the field amplitude to remain below the Jordan-frame reduced Planck scale, $\tilde{m}_{\mathrm{Pl}}^{2}(\phi)\equiv m^2_{\rm Pl}A(\phi)$. This ensures that the potential will be under control from quantum-gravity corrections\footnote{From an EFT perspective, one expects corrections to the potential of the form $\phi^n/m_\text{pl}^{n-4}$, so transplanckian field values would allow for non-negligible corrections at the Planck scale.}. This is an additional conservative modeling prior and is not required by the no-ghost conditions.

Finally, while the choice of coupling function is well motivated, we have chosen the simplest potential that has been previously studied in the context of early-time modifications of the standard cosmological model (e.g. Rock'n'Roll Early Dark Energy \cite{AGRAWAL2023101347} and Higgs-like modified gravity with $V(\phi)\propto \phi^4$) and that allows for a wide range of phenomenological behaviors. In this work, we focus our analysis on the cases $n=2,3,4$ and set aside the quadratic potential $n=1$ ($V(\phi) \propto \phi^2$). As established in \cite{Steinhardt:1994vs}, a quadratic potential undergoes rapid, high-frequency oscillations where the time-averaged scalar sector scales strictly as non-relativistic matter ($\rho_\phi \propto a^{-3}$). Consequently, $n=1$ cannot provide the stiff pre-recombination energy injection needed to alter the sound horizon and address the $H_0$ tension. Furthermore, tracking its dense oscillatory dynamics numerically within Boltzmann solvers without cycle-averaging approximations introduces integration instabilities, justifying our restriction to higher-power potentials.

\section{Effective Gravitational Constant and Observational Bounds}
\label{sec:constraints}

\begin{figure}
    \centering
    \includegraphics[width=1\linewidth]{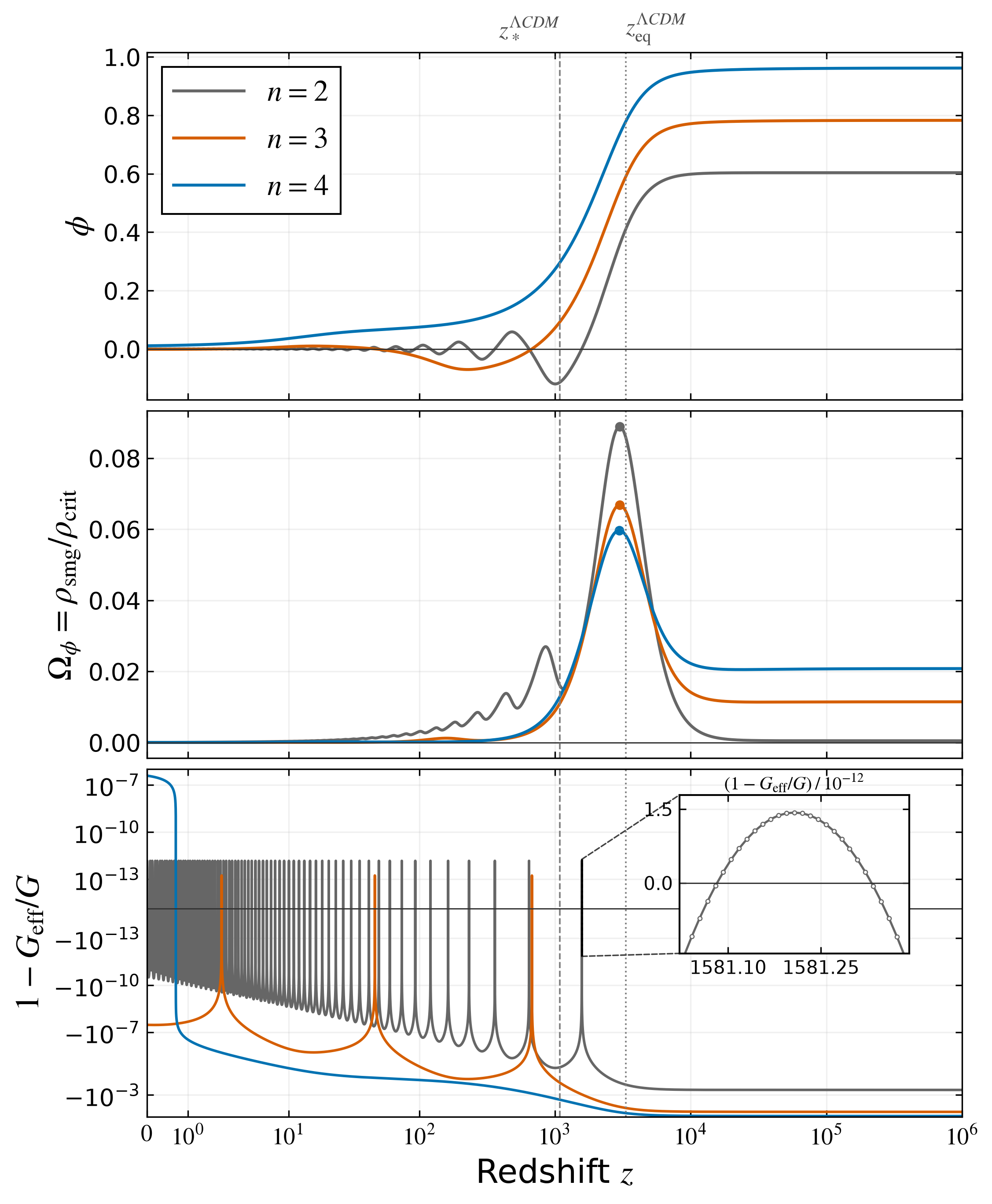}
    \caption{Scalar field evolution (upper panel), relative scalar-field energy density $\Omega_\phi\equiv\rho_{\rm smg}/\rho_{\rm crit}$ (middle panel), and deviation from unity of the effective Newton's gravitational constant (lower panel) for the P-LB-SN best-fits presented in table ~\ref{tab:constraints_P_PB}. We zoom-in into a peak to show that they are not divergence points, but instead very brief weaker gravity epochs: the horizontal axis is the redshift and the vertical axis is scaled by $(1-G_{\rm eff}/G)/10^{-12}$.}
    \label{fig:bestfit_scalar_background}
\end{figure}

Scalar-tensor theories in the form of eq.~\eqref{eq:original_action} produce a field-dependent Newton's Constant $G_\text{N} \equiv 1/8\pi A(\phi)$. This quantity represents the cosmological gravitational strength and is directly related to the expansion rate, but it is \textit{not} the quantity measured by Cavendish-type experiments~\cite{Uzan_2011}. Indeed, in the weak-field limit, the local gravitational attraction receives an additional contribution from the scalar exchange force. As a result, local experiments measure an effective Cavendish constant $G_\text{eff}$ instead, which for an effectively massless field or on sufficiently small scales ($k\rightarrow \infty$) is given by \cite{Gannouji_2006, Uzan_2011}
\begin{equation}
    \frac{G_\text{eff}}{G} = \frac{1}{A}\left(\frac{2A+4A_{,\phi}^2}{2A+3A_{,\phi}^2}\right).\label{eq:effective_grav}
\end{equation}
Then, $G_{\rm eff}$ is the relevant quantity for matter clustering, and the solutions we attempt to find should not only contain $G_\text{N}/G < 1$, but also $G_{\rm eff}/G < 1$ for growth to be suppressed. The third panel of fig.~\ref{fig:bestfit_scalar_background} exemplifies the evolution of $G_\text{eff}$ throughout the cosmic history for the three choices of potential. We note that the amplitude of the narrow peaks on the weaker-gravity regime for the $n=2,3$ cases is too short at late-times to affect large-scale observables; the specifics will be discussed in the forthcoming sections.

On the observational side, the present deviations of gravity from general relativity are quantified by the parametrized post-Newtonian (PPN) parameters, which for standard scalar-tensor theories are given as \cite{Boisseau_2000}
\begin{align}
    \gamma_\text{PPN} - 1 &= - \frac{A_{,\phi}^2}{A + 2A_{,\phi}^2},\label{eq:gamma_ppn}\\
    \beta_\text{PPN} - 1 &= \frac{AA_{,\phi}}{8A+12A_{,\phi}^2}\frac{d\gamma_\text{PPN}}{d\phi}.
\end{align}
Standard GR is characterized by $\gamma_\text{PPN} = \beta_\text{PPN} = 1$. The Shapiro delay measured with the Cassini spacecraft gives $\gamma_\text{PPN} - 1 = (2.1 \pm 2.3)\times 10^{-5}$~\cite{Bertotti_2003rm}, and planetary ephemerides including MESSENGER ranging to Mercury give $\beta_\text{PPN}-1 = (-4.1 \pm 7.8)\times 10^{-5}$~\cite{Will_2014}, both at 68\% confidence level. Note that since the right-hand side of eq.~\eqref{eq:gamma_ppn} is never positive in our theory, only the lower Cassini limit is relevant, and it is equivalent to
\begin{equation}
    \frac{A_{,\phi}^2}{A}\bigg|_{\rm today} < 2.0\times 10^{-6}.\label{eq:upper_bound_cassini}
\end{equation}

This bound places a direct constraint on the linear coefficient $c_1$. Noting that the quadratic coupling can be cast as $A_{,\phi}^2 = c_1^2 + 4c_2(A-1)$, and working under the expectation that the field today sits at the origin, $A(\phi_0)=1$, eq.~\eqref{eq:upper_bound_cassini} gives $|c_1| < 1.4\times10^{-3}$. Larger values of $|c_1|$ remain allowed if the field today lies close to $\phi_\star=-c_1/(2c_2)$ instead, as at this point we would also have $A_{,\phi}(\phi_\star)=0$.

Furthermore, the present-day time variation of $G_\text{eff}$ is constrained by Lunar Laser Ranging measurements to the value \cite{universe7020034}
\begin{equation}
\left.
\frac{\dot{G}_{\rm eff}}{G_{\rm eff}}
\right|_{z=0}
=
(-5.0\pm9.6)\times10^{-15}\ {\rm yr}^{-1}.
\end{equation}

Finally, Big Bang Nucleosynthesis (BBN) can be used to constrain the value of $G_\text{N}(\phi_{\rm bbn})/G = 1/A(\phi_{\rm bbn})$. This quantity is related to the expansion rate during the radiation era, impacting the freeze-out of neutron-proton interactions and the subsequent primordial light element abundances. Here we use the $3\sigma$ BBN bound on the gravitational constant \cite{Alvey_2020}
\begin{equation}
    \frac{G_\text{N}(t_\text{BBN})}{G} \in [0.895, 1.08],
\end{equation}
and also consider the effects of the MG sector on the effective number of neutrinos, as was done in \cite{Zumalac_rregui_2020}.

With these constraints, we can estimate the maximum amplitude of the weaker-gravity regime. First, eq.~\eqref{eq:effective_grav} implies $G_{\rm eff}/G\geq1/A$, so that $A>1$ is necessary for $G_{\rm eff}<G$. For $c_2<0$, the coupling function is an inverted parabola with global maximum at $\phi_\star=-c_1/(2c_2)$, where $A_{,\phi}(\phi_\star)=0$, and therefore
\begin{equation}
    A(\phi)\leq A(\phi_\star) = 1+\frac{c_1^2}{4|c_2|}
    \Longrightarrow
    \frac{G_{\rm eff}}{G}\geq\frac{1}{1+\frac{c_1^2}{4|c_2|}},\label{eq:lower_Geff_bound}
\end{equation}
independently of whether the field ever reaches $\phi_\star$.

A significant weaker-gravity regime thus requires $|c_2|\lesssim c_1^2$, i.e. $|c_2|\lesssim10^{-6}$ for $|c_1|\lesssim10^{-3}$, but the same coefficient sets the early enhancement of the gravitational coupling. Defining $\varepsilon\equiv1-A(\phi_{\rm ini})$, eq.~\eqref{eq:coupling} gives $|c_2|=(\varepsilon+c_1\phi_{\rm ini})/\phi_{\rm ini}^2$, so that
\begin{equation}
    1-\frac{G_{\rm eff}}{G}\leq\frac{c_1^2\phi_{\rm ini}^2}{4\left(\varepsilon-|c_1|\phi_{\rm ini}\right)}.
    \label{eq:minimum_Geff}
\end{equation}
For a percent-level early enhancement, $\varepsilon\gtrsim10^{-2}$, and $|c_1|\phi_{\rm ini}\lesssim10^{-3}$, the weaker-gravity regime is limited to $1-G_{\rm eff}/G\lesssim10^{-4}$. A deeper weaker phase requires $\varepsilon$ to be tuned close to $|c_1|\phi_{\rm ini}$, so that the early phase becomes negligible; the linear term then dominates the coupling and the positivity of the energy density at the crossings of the origin, eq.~\eqref{eq:constraint_c1}, suppresses $c_1$ further. The early stronger-gravity and late weaker-gravity phases therefore compete for the same coefficients.

These estimates assume that the field sits at the origin today, when evaluating the Cassini bound, and order-of-magnitude values of $\varepsilon$ and $\phi_{\rm ini}$. The actual boundaries of the viable region depend on the full background evolution, through the energy conditions of eq.~\eqref{eq:condition_general} and the epoch at which the field thaws, and must be determined numerically. We obtain them in the next section through a systematic exploration of the parameter space.

\section{Parameter Space Exploration}
\label{sec:exploration}

To systematically probe the regions where background solutions exist and satisfy all the conditions outlined in the previous sections, we initially applied a Sobol Sequence method \cite{sobol1967distribution} to sample the modified gravity parameters across a wide domain for a total of $2^{21}$ points. We then solved the background equations using a modified version of the Boltzmann solver \texttt{hi\_class}, which retains only cosmological histories satisfying the theoretical and observational bounds described in sections \ref{sec:theory} and \ref{sec:constraints}. In addition, since we desire to study the prior space that contains a stronger-to-weaker gravity transition, we select the subclass of histories that contains a weaker-gravity regime by demanding that $G_{\rm eff}(\phi)<G$ at least once during the cosmological evolution. No minimum amplitude or duration of this regime is imposed in this study\footnote{This allows solutions in which $G_{\rm eff}<G$ only briefly or by a very small amount. We therefore select solutions in which gravity is not always stronger or equal than its present bare value, and do not assume a theoretically preferred strength or duration of the weaker-gravity phase.}.

The use of eq.~\eqref{eq:effective_grav} for this test does not guarantee that $G_{\rm eff}$ is weaker on non-quasistatic regimes or at all scales, but it serves as a reference quantity for modes that are deep within the Hubble horizon, which are known to have significant effects on the formation of structure \cite{DEFELICE2011123, Sawicki_2015}. In fact, the weaker-gravity transitions found in our solutions occur close to
$\phi=0$, where the potential and its first and second derivatives vanish for $n = 2,3,4$. In this region, the effective mass of the scalar field has only the curvature-dependent contribution \cite{DEFELICE2011123}
\begin{equation}
    m_{\rm eff,bg}^2|_{\phi\simeq 0} = \left(V_{,\phi\phi} -\frac12A_{,\phi\phi}R\right)_{\phi\simeq 0} \simeq -c_2R,
\end{equation}
for $c_2 < 0$, as will be found later. Then,
\begin{equation}
\left.m_{\rm eff,bg}^{2}\right|_{\phi\simeq0}
\simeq
\begin{cases}
0, & \text{rad. dom.},\\
\mathcal{O}(H^2) & \text{otherwise}
\end{cases}
\label{eq:cases_eff_mass}
\end{equation}
Thus, for modes well inside the Hubble radius, $k^2/a^2\gg H^2$, one also expects $k^2/a^2\gg m_{\rm eff,bg}^2$, so that the massless expression for $G_{\rm eff}$ is a good approximation to check the sign change of $G_{\rm eff} - G$.

Given the analytical relation between $G_{\rm eff}$ and the coupling function, the above conditions may be understood as shaping the prior probability range of the free parameters, as the initial conditions determine the scalar field evolution. Furthermore, this choice also prevents the theory from reducing to the usual EMG scenarios with $c_1 = 0$, and so we explore a more diverse, non-trivial region of the parameter space to search for novel mechanisms. Nevertheless, since satisfying the prior conditions depends on the background evolution of the scalar field and matter content, there is no simple, exact form of restricting the modified gravity parameters from this requirement alone.

\begin{figure}
    \centering
    \includegraphics[width=1\linewidth]{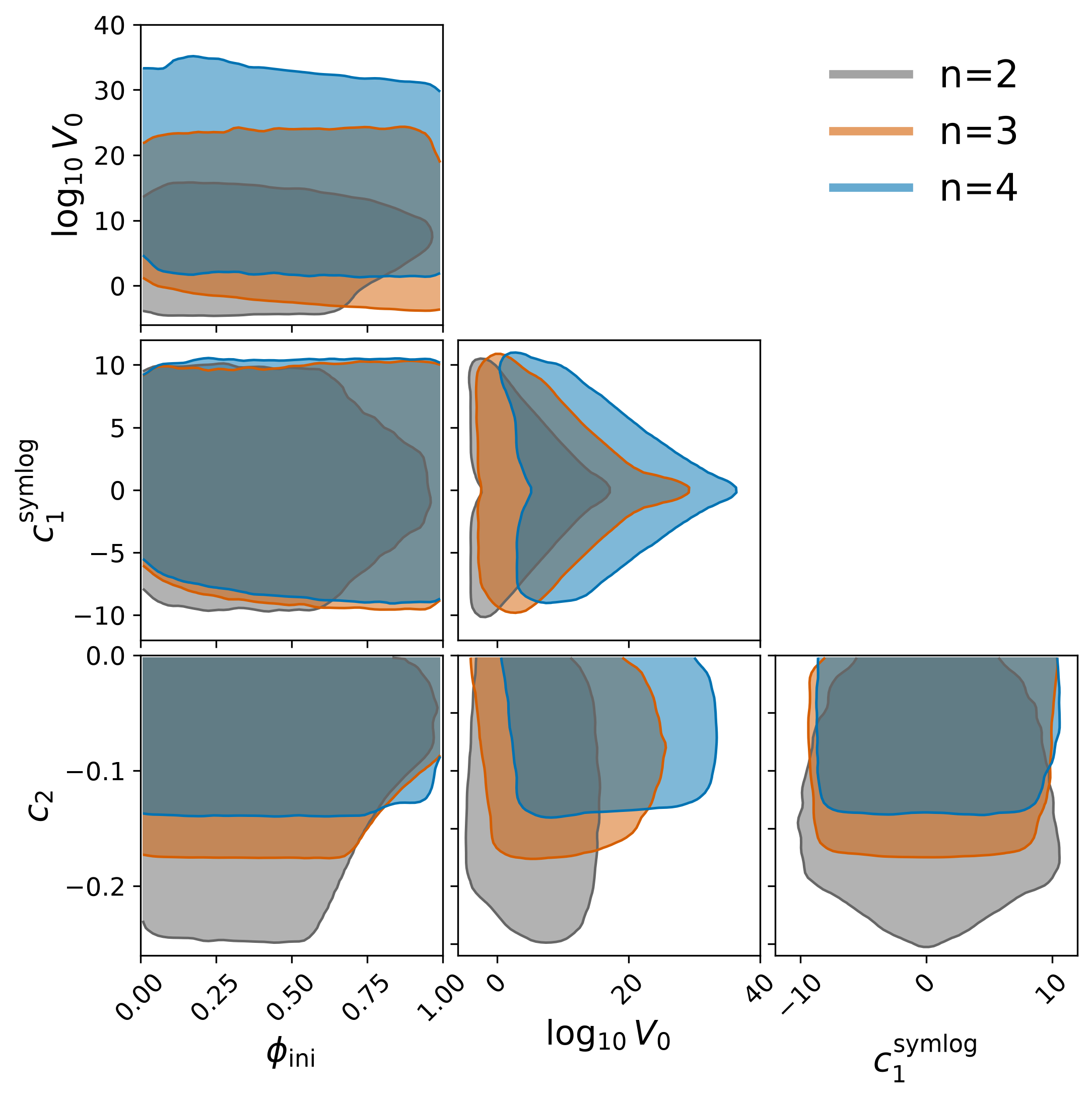}
    \caption{Triangle plot showing the regions of the modified gravity parameter space that contain background solutions after imposing all observational and theoretical bounds presented in Secs.~\ref{sec:theory} and~\ref{sec:constraints}. Note that $c_1^\text{symlog}\in[-11,11]$ implies $c_1 \in [-0.003, 0.003]$.}
    \label{fig:validity_regions}
\end{figure}

The results of the parameter space exploration are presented in fig.~\ref{fig:validity_regions}. It was found that the $c_1$ parameter does not have a sign restriction and spans several orders of magnitude, and so we sampled the transformed variable $c_1^\text{symlog}$, defined such that $c_1 \equiv 10^{-7} \sinh(c_1^\text{symlog})$. This allows the sampler to efficiently explore values exponentially close to zero while maintaining symmetry around the origin.

Furthermore, we found that the vast majority of points in the viability regions contained an oscillatory behavior and a scalar field crossing zero; this is presented in table~\ref{tab:zero_crossings}. This result emphasizes that restricting a model to only non-oscillatory solutions, as is standard in the Effective Theory of Dark Energy \cite{GiuliaGubitosi_2013, FRUSCIANTE20201}, may ignore a large, viable region of the parameter space that has interesting physics. 
\begin{table}[ht]
    \centering
    \renewcommand{\arraystretch}{1.3}
    \begin{tabular}{l|ccc}
        \hline\hline
        Scalar Field Behavior & $n=2$ & $n=3$ & $n=4$ \\
        \hline
        Oscillatory & 95\% & 90\% & 96\% \\
        Crosses Zero & 99\% & 98\% & 96\% \\
        \hline\hline
    \end{tabular}
    \caption{Percentage of viable points that exhibit oscillatory behavior in the past ($z>0$) and total zero-crossings for each potential power $n$.}
    \label{tab:zero_crossings}
\end{table}

\begin{table*}
    \centering
    \renewcommand{\arraystretch}{1.4}
    \begin{tabular}{llccc}
        \hline
        Likelihood & Data & P-L & P-LB & P-LB-SN \\
        \hline
        \texttt{planck\_2018\_lowl.EE} & Low-$\ell$ E-mode polarization & \checkmark & \checkmark &  \checkmark \\
        \texttt{planck\_2018\_lowl.TT} & Low-$\ell$ temperature & \checkmark & \checkmark & \checkmark \\
        \texttt{planck\_2018\_highl\_plik.TTTEEE\_lite} & High-$\ell$ TT, TE, EE (lite) & \checkmark & \checkmark & \checkmark\\
        \texttt{planckpr4lensing} & Planck PR4 CMB lensing & \checkmark & \checkmark &  \checkmark\\
        \texttt{bao.desi\_dr2.desi\_bao\_all} & DESI DR2 & & \checkmark & \checkmark \\
        \texttt{sn.pantheonplusshoes} & Type Ia Supernovae + SH0ES Calibration & & & \checkmark \\
        \hline
    \end{tabular}
    \caption{Likelihoods and dataset combinations used in the nested sampling runs.}
    \label{tab:likelihoods_runs}
\end{table*}

Since the scalar field crosses zero in more than $96\%$ of the points found, the constraint on $c_1$ from eq.~(\ref{eq:constraint_c1}) holds in practically all of the parameter space. This finding explains the funneling observed in the $c_1^\text{symlog}\times \log_{10}V_0$ contour plot of fig.~\ref{fig:validity_regions}: as the potential amplitude increases, the field unfreezes sooner and the Hubble parameter pushes $c_1$ to values close to zero. On the opposite direction, a lower potential amplitude allows the field to unfreeze when $H(t)$ is lower and so most of the constraint on $c_1$ comes from the Cassini bound on the Shapiro delay. Nevertheless, $V_0$ cannot be too low, otherwise the field would unfreeze too late and current bounds on modified gravity parameters would be violated.

Furthermore, the $c_2$ parameter must be negative after applying our theoretical bounds. We found that relaxing the weaker gravity condition allows $c_2$ to reach values $\sim 0.01$ in all scenarios, which is not a substantial increase in the allowed parameter space. In fact, the positivity of the energy density is the most restrictive condition, and it prevents $c_2$ from having higher values. This was also observed in ref.~\cite{Braglia:2020auw}, where solutions with positive quadratic coupling $\xi \equiv c_2 \geq 0.1$ were shown to have a negative density parameter in the early Universe.

Regarding $1-G_{\rm eff}/G$, the Sobol exploration finds a maximum depth of $(1-G_{\rm eff}/G)_{\rm max} = 2.01\times 10^{-4}$ across all potential choices. This confirms the rough estimate made in the previous section and shows that the weaker gravity regimes will be shallow. Nevertheless, no immediate restriction could be placed on their duration or epoch of occurrence, as Solar System and Lunar Laser Ranging tests limit the coupling only today, and BBN only at $t_{\rm BBN}$; between these epochs, the history of $G_{\rm eff}/G$ is constrained solely by cosmological data. The latter constraint, however, is indirect. Since the weaker-gravity excursions happen after the thawing of the scalar field, which in turn affects the expansion rate and may alter the physics before/during recombination, the epoch and duration for which $G_{\rm eff}/G<1$ are constrained by CMB, BAO and supernovae data. Whether the data favors solutions with brief, frequent, and/or long weaker phases, and whether it can coexist with the early energy injection that reduces the sound horizon, can therefore only be established by a statistical analysis which we pursue in the next Section. In addition, the data will also give insight into whether the weaker phase can affect structure formation at all, since the growth of matter perturbations responds to the accumulated effects of the weaker regime rather than to its maximum depth alone.

To address these topics, we confront the theory with the Planck 2018 temperature and polarization dataset \cite{Planck_2020_likelihoods}, the PR4 lensing package \cite{Carron_2022}, the Data Release 2 from the Dark Energy Spectroscopic Instrument (DESI DR2) \cite{DESI:2025zgx} and the supernova data from the Pantheon$+$ plus the SH0ES galaxy-distance calibration using Cepheids \cite{Brout:2022vxf}. The statistical analysis is carried out with the PolyChord sampler \cite{Handley_2015} via the \texttt{Cobaya} framework \cite{Torrado_2021, 2019ascl_soft10019T}, as its nested sampling algorithm is well-suited to handle complex, multi-modal posterior distributions. We used an evidence tolerance of 0.1, 275 live points and 22 repeats in all cases. The best-fit values are found via the Py-Bobyqa minimizer algorithm \cite{Cartis_2018, Cartis_2021}, and the plots are then handled by the GetDist package \cite{Lewis_2025}; the likelihoods used in each run are indicated in table~\ref{tab:likelihoods_runs}. For the computation of non-linear matter perturbations, we employ the \textit{halofit} model \cite{Smith_2003, Takahashi_2012}. The full input configuration for the \texttt{hi\_class} code is presented in Appendix~\ref{app:hi_class}.

The final flat priors adopted for our statistical analyses are divided into standard cosmological parameters and modified gravity parameters. The standard $\Lambda$CDM parameters are assigned flat, uninformative priors common to all models, as detailed in table~\ref{tab:lcdm_priors}. For the modified gravity sector, we fix the initial field derivative $\phi'_\text{ini} = 0$ and sample the remaining parameters ($\phi_\text{ini}$, $\log_{10} V_0$, $c_1^\text{symlog}$, and $c_2$) based on the initial Sobol exploration\footnote{Note that we are focusing on the parameter space that has weaker-gravity epochs.}. The flat prior ranges are summarized in table~\ref{tab:mg_priors}.

\begin{table}
    \centering
    \renewcommand{\arraystretch}{1.3}
    \begin{tabular}{l|c}
        \hline\hline
        Parameter & Prior \\
        \hline
        $\omega_b$ & $[0.013, 0.033]$ \\
        $\omega_\text{cdm}$ & $[0.001, 0.99]$ \\
        $H_0$ & $[40, 100]$ \\
        $\tau_\text{reio}$ & $[0.01, 0.8]$ \\
        $\ln(10^{10}A_s)$ & $[1.7, 5.0]$ \\
        $n_s$ & $[0.7, 1.3]$ \\
        \hline\hline
    \end{tabular}
    \caption{Flat priors for the standard $\Lambda$CDM cosmological parameters adopted across all models in our statistical analyses.}
    \label{tab:lcdm_priors}
\end{table}

\begin{table}
    \centering
    \renewcommand{\arraystretch}{1.3}
    \begin{tabular}{l|c|c|c}
        \hline\hline
        Parameter & $n=2$ & $n=3$ & $n=4$ \\
        \hline
        $\phi_\text{ini}$ & $[0, 1]$ & $[0, 1]$ & $[0, 1]$ \\
        $\log_{10} V_0$ & $[-6, 16]$ & $[-6, 31]$ & $[0, 40]$ \\
        $c_1^\text{symlog}$ & $[-11, 11]$ & $[-11, 11]$ & $[-11, 11]$ \\
        $c_2$ & $[-0.26, 0.0]$ & $[-0.18, 0.0]$ & $[-0.15, 0.0]$ \\
        \hline\hline
    \end{tabular}
    \caption{Flat prior ranges for the modified gravity parameters used in the statistical analysis for the quartic ($n=2$), sextic ($n=3$), and octic ($n=4$) potential cases. Note that the upper limit on the initial scalar field value is ultimately overwritten by the sub-planckian field condition $\phi_{\rm ini}^2 <A(\phi_{\rm ini})$.}
    \label{tab:mg_priors}
\end{table}

The choice to set $\phi'_\text{ini} = 0$ is motivated by the expectation that the field is frozen at very early times due to the high Hubble friction, and indeed this is seen in all solutions found. However, one may attempt to connect the initial conditions of the scalar sector at the radiation dominated epoch to some inflationary+reheating theory, which may then motivate a non-zero initial field velocity. The exploration of this path is left to a future work.

\section{Results}

We now present the cosmological constraints on the three power-law potentials, $V(\phi)\propto V_0\phi^{2n}$ with $n=2,3,4$.  We denote by P-L the combination of Planck temperature and polarization data with the Planck PR4 lensing, by P-LB the addition of DESI DR2 BAO, and by P-LB-SN the further addition of Pantheon+ together with the SH0ES calibration. Unless explicitly stated otherwise, the constraints are obtained via the marginalized posterior means and $68\%$ credible intervals.

\begin{figure*}
    \centering
    \includegraphics[width=1\linewidth]{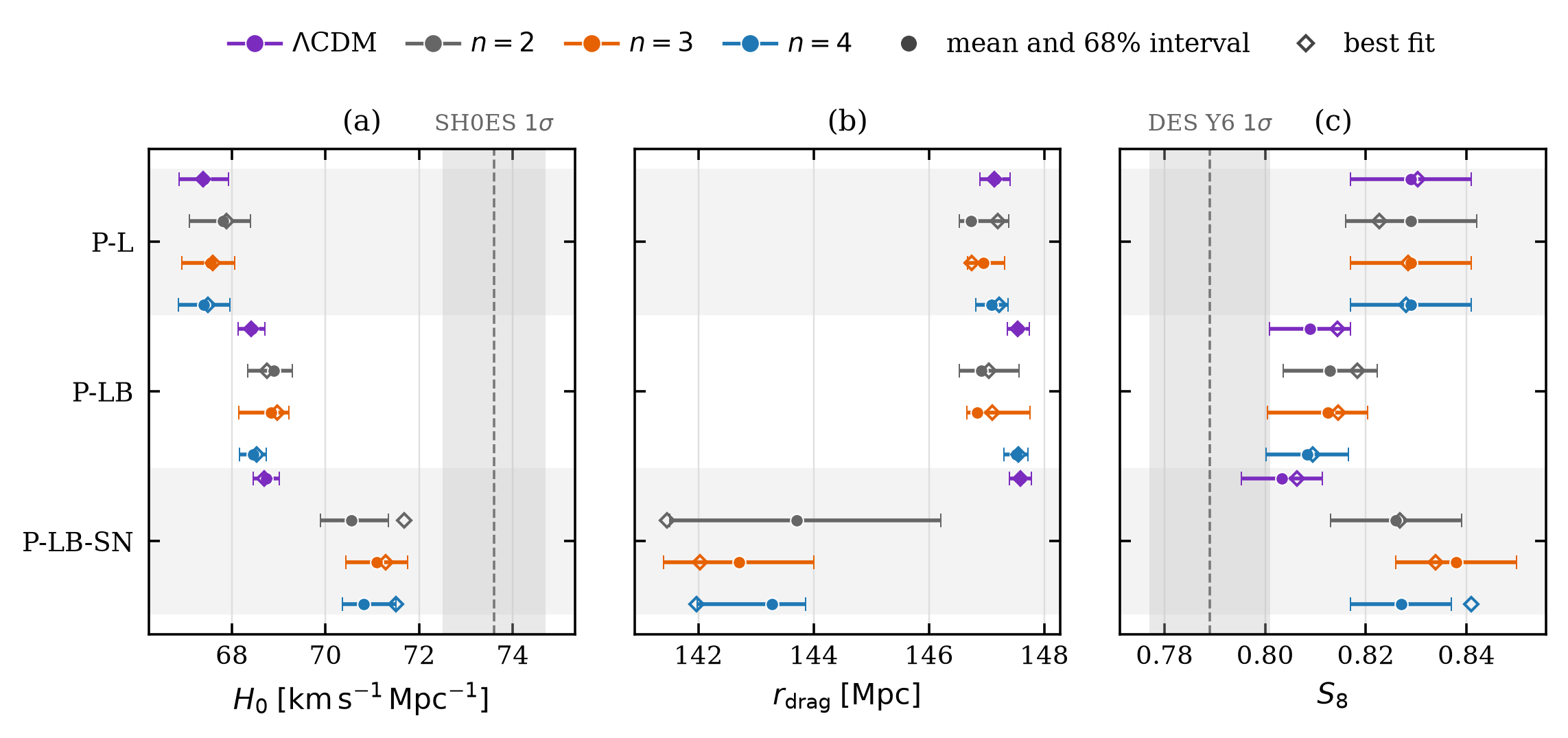}
    \caption{Overview of resulting constraints on the $H_0$, $r_{\rm drag}$ and $S_8$ parameters for the P-L, P-LB, and P-LB-SN datasets and the three exponents of the power-law potential $n=2,3,4$. The dots and bars correspond to the mean values and $1\sigma$ confidence level, while the diamonds mark the best-fit values. The gray vertical bands in the $H_0$ and $S_8$ panels indicates the late-time measurements $H_0 = 73.6 \pm 1.1$ km/s/Mpc from the Pantheon$+$ and SH0ES combined analysis \cite{Brout:2022vxf} and $S_8 = 0.789\pm 0.012$ from the DES Year 6 galaxy clustering and weak lensing ($3\times2$pt) analysis \cite{DESY6}.}
    \label{fig:overview}
\end{figure*}

The marginalized constraints and best-fit values are collected in table~\ref{tab:constraints_P_PB}, and a visual overview of all datasets and models constraints on the $H_0$, $r_{\rm drag}$ and $S_8$ parameters is presented in Fig.~\ref{fig:overview}. The $\chi^2$ and evidence $\mathcal{Z}$ analyses of each dataset are presented in section~\ref{subsec:model_comparison}.

\subsection{CMB-only and CMB+BAO datasets}
\label{subsec:results-p-pb}

For the P-L dataset, the six standard cosmological parameters are remarkably stable against the extension of the gravitational sector, and the inferred Hubble constant remains consistent with the base-$\Lambda$CDM model. The modified-gravity parameters are only weakly constrained, and the scalar sector is suppressed in two different ways. For $n=3$ and $n=4$, the potential amplitude is driven to large values, $\log_{10}V_0\simeq11$ and $17$, so that the field thaws and decays well before recombination. For $n=2$, the amplitude is lower, $\log_{10}V_0=2.4^{+3.0}_{-4.4}$, but the posterior favors small initial field values, $\phi_{\rm ini}=0.175^{+0.062}_{-0.15}$, which keep both the early enhancement of the gravitational coupling, $G_\text{N}(\phi_{\rm ini})/G\simeq1.005$, and the initial potential energy, $V(\phi_{\rm ini})\propto\phi_{\rm ini}^4$, small; the $n=2$ best fit lies instead in the high-amplitude mode, $\log_{10}V_0\simeq9.1$, discussed in Sec.~\ref{subsec:results-mg}. Therefore, with CMB data alone, the additional scalar-tensor freedom is mostly suppressed and the effective history is that of $\Lambda$CDM.

\begin{figure}[H]
    \centering
    \includegraphics[width=1.0\linewidth]{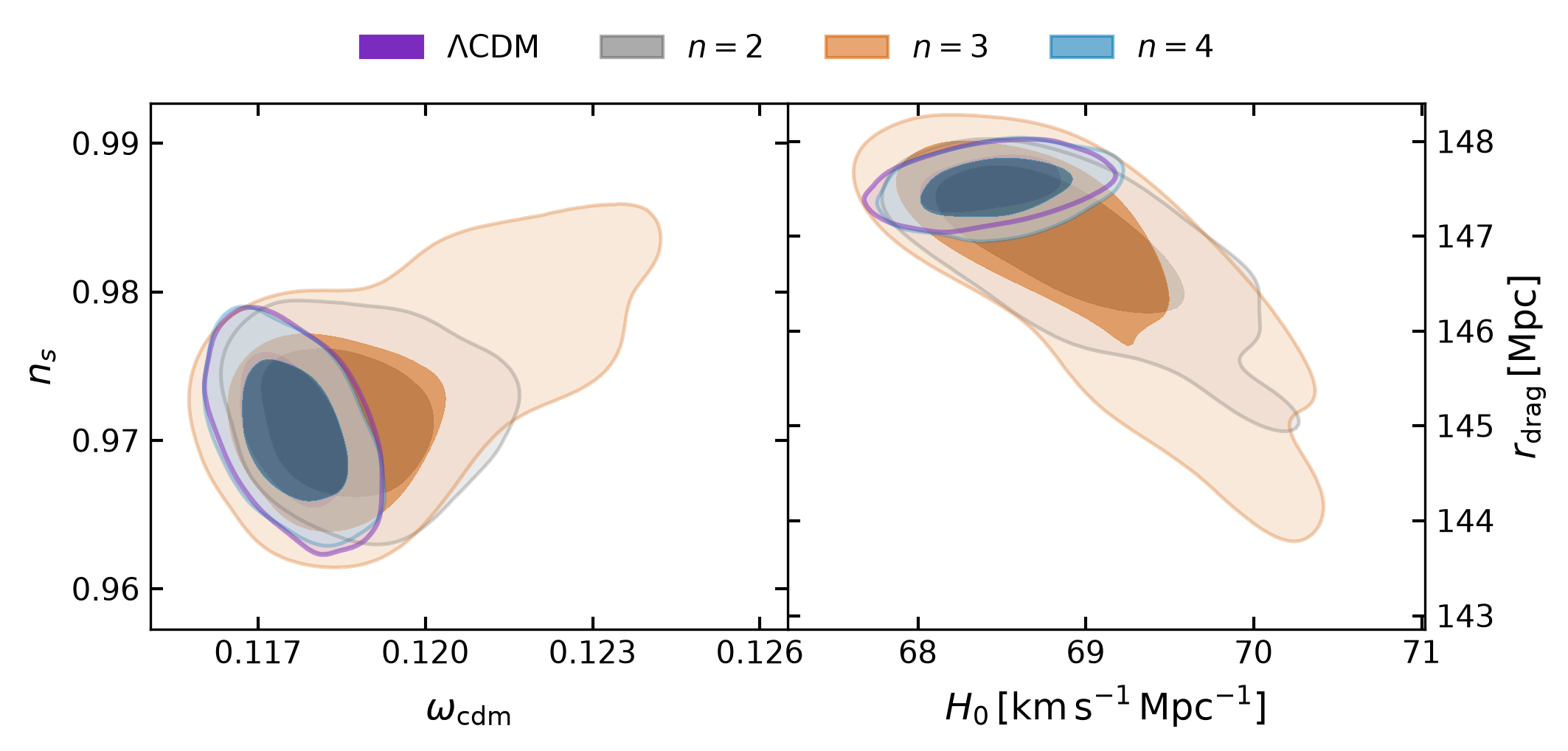}
    \caption{Contour plots of the $H_0\times r_{\rm drag}$ and $n_s\times\omega_{\rm cdm}$ parameters for the P-LB dataset.}
    \label{fig:non-gaussian_tail}
\end{figure}

Adding the DESI DR2 BAO dataset partially breaks this behavior and a lower potential is preferred, allowing the scalar field to remain relevant until later times. The $n=2$ and $n=3$ posteriors present a pronounced non-Gaussian tail toward larger $H_0$ and smaller $r_{\rm drag}$, whereas the $n=4$ posterior remains much more localized and close to the $\Lambda$CDM region. This difference is also visible in the standard parameters: P-LB allows a broader high-$\omega_{\rm{cdm}}$, high-$n_s$ tail for $n=2,3$, while $n=4$ is more tightly constrained. These results are shown in fig.~\ref{fig:non-gaussian_tail}.

\begin{table*}
\centering
\scriptsize
\setlength{\tabcolsep}{6.5pt}
\renewcommand{\arraystretch}{0.96}
\resizebox{\textwidth}{!}{%
\begin{tabular}{lcccc}
\hline\hline
\multicolumn{5}{c}{P-L}\\
\hline
Parameter
& \(\boldsymbol{\Lambda\mathrm{CDM}}\)
& \(\boldsymbol{n=2}\)
& \(\boldsymbol{n=3}\)
& \(\boldsymbol{n=4}\)\\
\hline

\multicolumn{5}{l}{\textit{Sampled cosmological parameters}}\\
\(100\omega_b\)
& \margbf{2.237\pm0.015}{2.236}
& \margbf{2.240\pm0.016}{2.244}
& \margbf{2.238\pm0.015}{2.240}
& \margbf{2.237\pm0.015}{2.235}\\
\(\omega_{\mathrm{cdm}}\)
& \margbf{0.1199\pm0.0012}{0.1199}
& \margbf{0.1204^{+0.0012}_{-0.0017}}{0.1192}
& \margbf{0.1201\pm0.0012}{0.1205}
& \margbf{0.1199\pm0.0012}{0.1196}\\
\(H_0\)
& \margbf{67.40\pm0.53}{67.38}
& \margbf{67.81^{+0.59}_{-0.72}}{67.89}
& \margbf{67.55^{+0.51}_{-0.62}}{67.59}
& \margbf{67.40\pm0.55}{67.48}\\
\(\tau\)
& \margbf{0.0538\pm0.0074}{0.0552}
& \margbf{0.0542\pm0.0077}{0.0591}
& \margbf{0.0542\pm0.0074}{0.0498}
& \margbf{0.0539\pm0.0075}{0.0524}\\
\(\ln(10^{10}A_{\mathrm{s}})\)
& \margbf{3.042\pm0.014}{3.044}
& \margbf{3.045\pm0.015}{3.051}
& \margbf{3.043\pm0.014}{3.036}
& \margbf{3.042\pm0.014}{3.044}\\
\(n_{\mathrm{s}}\)
& \margbf{0.9652\pm0.0040}{0.9652}
& \margbf{0.9665\pm0.0044}{0.9676}
& \margbf{0.9652\pm0.0043}{0.9659}
& \margbf{0.9648\pm0.0042}{0.9652}\\
\multicolumn{5}{l}{\textit{Sampled modified-gravity parameters}}\\
\(\phi_{\mathrm{ini}}\)
& \notapp
& \margbf{0.175^{+0.062}_{-0.15}}{0.086}
& \margbf{0.36^{+0.10}_{-0.34}}{0.22}
& \margbf{0.50^{+0.31}_{-0.28}}{0.43}\\
\(\log_{10}V_0\)
& \notapp
& \margbf{2.4^{+3.0}_{-4.4}}{9.1}
& \margbf{11.0^{+6.1}_{-5.5}}{5.6}
& \margbf{16.6^{+3.8}_{-6.5}}{17.6}\\
\(c_2\)
& \notapp
& \margbf{-0.113\pm0.059}{-0.014}
& \margbf{-0.078^{+0.071}_{-0.030}}{-0.050}
& \margbf{-0.062^{+0.043}_{-0.034}}{-0.024}\\
\multicolumn{5}{l}{\textit{Derived parameters}}\\
\(\Omega_{\mathrm{m}}\)
& \margbf{0.3146\pm0.0073}{0.3147}
& \margbf{0.3120\pm0.0080}{0.3086}
& \margbf{0.3138\pm0.0076}{0.3143}
& \margbf{0.3148\pm0.0075}{0.3132}\\
\(\Omega_{\Lambda}\)
& \margbf{0.6853\pm0.0073}{0.6852}
& \margbf{0.6879\pm0.0080}{0.6913}
& \margbf{0.6861\pm0.0076}{0.6856}
& \margbf{0.6851\pm0.0075}{0.6867}\\
\(100\omega_{\rm m}\)
& \margbf{14.29\pm0.11}{14.29}
& \margbf{14.34^{+0.11}_{-0.17}}{14.22}
& \margbf{14.31\pm0.12}{14.36}
& \margbf{14.30\pm0.11}{14.26}\\
\(\sigma_8\)
& \margbf{0.8097\pm0.0056}{0.8106}
& \margbf{0.8130^{+0.0063}_{-0.0079}}{0.8111}
& \margbf{0.8105^{+0.0057}_{-0.0067}}{0.8093}
& \margbf{0.8097\pm0.0057}{0.8100}\\
\(r_{\mathrm{drag}}\,[\mathrm{Mpc}]\)
& \margbf{147.14\pm0.26}{147.14}
& \margbf{146.73^{+0.65}_{-0.21}}{147.20}
& \margbf{146.94^{+0.37}_{-0.27}}{146.74}
& \margbf{147.09\pm0.28}{147.22}\\
\(G(\phi_{\mathrm{ini}})/G_0\)
& \margbf{1}{1}
& \margbf{1.00497^{+0.00082}_{-0.0055}}{1.000100}
& \margbf{1.01301^{+0.00096}_{-0.015}}{1.00235}
& \margbf{1.0195^{+0.0068}_{-0.020}}{1.0044}\\
\(S_8\)
& \margbf{0.829\pm0.012}{0.830}
& \margbf{0.829\pm0.013}{0.823}
& \margbf{0.829\pm0.012}{0.828}
& \margbf{0.829\pm0.012}{0.828}\\
\hline\hline
\multicolumn{5}{c}{P-LB}\\
\hline
Parameter
& \(\boldsymbol{\Lambda\mathrm{CDM}}\)
& \(\boldsymbol{n=2}\)
& \(\boldsymbol{n=3}\)
& \(\boldsymbol{n=4}\)\\
\hline

\multicolumn{5}{l}{\textit{Sampled cosmological parameters}}\\
\(100\omega_b\)
& \margbf{2.253\pm0.013}{2.252}
& \margbf{2.251\pm0.014}{2.243}
& \margbf{2.257\pm0.017}{2.249}
& \margbf{2.255\pm0.013}{2.259}\\
\(\omega_{\mathrm{cdm}}\)
& \margbf{0.11763\pm0.00065}{0.11769}
& \margbf{0.11869^{+0.00087}_{-0.0012}}{0.11865}
& \margbf{0.11882^{+0.00057}_{-0.0018}}{0.11814}
& \margbf{0.11763\pm0.00065}{0.11737}\\
\(H_0\)
& \margbf{68.42\pm0.29}{68.41}
& \margbf{68.89^{+0.40}_{-0.56}}{68.75}
& \margbf{68.84^{+0.37}_{-0.70}}{68.97}
& \margbf{68.46^{+0.27}_{-0.30}}{68.53}\\
\(\tau\)
& \margbf{0.0611\pm0.0071}{0.0575}
& \margbf{0.0586^{+0.0066}_{-0.0076}}{0.0557}
& \margbf{0.0599\pm0.0077}{0.0601}
& \margbf{0.0614\pm0.0073}{0.0617}\\
\(\ln(10^{10}A_{\mathrm{s}})\)
& \margbf{3.053\pm0.014}{3.069}
& \margbf{3.051\pm0.014}{3.054}
& \margbf{3.053^{+0.014}_{-0.017}}{3.062}
& \margbf{3.054\pm0.015}{3.061}\\
\(n_{\mathrm{s}}\)
& \margbf{0.9707\pm0.0034}{0.9668}
& \margbf{0.9715\pm0.0034}{0.9715}
& \margbf{0.9721^{+0.0034}_{-0.0053}}{0.9718}
& \margbf{0.9707\pm0.0032}{0.9724}\\
\multicolumn{5}{l}{\textit{Sampled modified-gravity parameters}}\\
\(\phi_{\mathrm{ini}}\)
& \notapp
& \margbf{0.23\pm0.11}{0.23}
& \margbf{0.41^{+0.17}_{-0.26}}{0.25}
& \margbf{0.47^{+0.25}_{-0.37}}{0.21}\\
\(\log_{10}V_0\)
& \notapp
& \margbf{1.3^{+1.5}_{-4.0}}{-2.1}
& \margbf{8.5^{+6.0}_{-8.0}}{2.3}
& \margbf{16.1^{+4.1}_{-6.1}}{18.8}\\
\(c_2\)
& \notapp
& \margbf{-0.131^{+0.038}_{-0.066}}{-0.178}
& \margbf{-0.079^{+0.060}_{-0.038}}{-0.133}
& \margbf{>-0.0796}{-0.110}\\
\multicolumn{5}{l}{\textit{Derived parameters}}\\
\(\Omega_{\mathrm{m}}\)
& \margbf{0.3008\pm0.0037}{0.3010}
& \margbf{0.2989\pm0.0040}{0.2998}
& \margbf{0.2997^{+0.0043}_{-0.0038}}{0.2970}
& \margbf{0.3005\pm0.0037}{0.2994}\\
\(\Omega_{\Lambda}\)
& \margbf{0.6991\pm0.0037}{0.6989}
& \margbf{0.7011\pm0.0040}{0.7001}
& \margbf{0.7002^{+0.0038}_{-0.0043}}{0.7029}
& \margbf{0.6995\pm0.0037}{0.7006}\\
\(100\omega_{\rm m}\)
& \margbf{14.081\pm0.063}{14.086}
& \margbf{14.184^{+0.081}_{-0.12}}{14.172}
& \margbf{14.203^{+0.050}_{-0.18}}{14.128}
& \margbf{14.082\pm0.063}{14.060}\\
\(\sigma_8\)
& \margbf{0.8079\pm0.0058}{0.8130}
& \margbf{0.8145^{+0.0074}_{-0.010}}{0.8185}
& \margbf{0.8128^{+0.0061}_{-0.012}}{0.8186}
& \margbf{0.8077\pm0.0060}{0.8102}\\
\(r_{\mathrm{drag}}\,[\mathrm{Mpc}]\)
& \margbf{147.55\pm0.19}{147.54}
& \margbf{146.91^{+0.65}_{-0.38}}{147.04}
& \margbf{146.83^{+0.92}_{-0.18}}{147.10}
& \margbf{147.51\pm0.21}{147.55}\\
\(G(\phi_{\mathrm{ini}})/G_0\)
& \margbf{1}{1}
& \margbf{1.0091^{+0.0078}_{-0.0094}}{1.0097}
& \margbf{1.0151^{+0.0046}_{-0.016}}{1.0083}
& \margbf{1.0173^{+0.0057}_{-0.020}}{1.0049}\\
\(S_8\)
& \margbf{0.8089\pm0.0081}{0.8144}
& \margbf{0.8129\pm0.0093}{0.8183}
& \margbf{0.8124^{+0.0080}_{-0.012}}{0.8145}
& \margbf{0.8083\pm0.0082}{0.8094}\\
\hline\hline
\multicolumn{5}{c}{P-LB-SN}\\
\hline
Parameter
& \(\boldsymbol{\Lambda\mathrm{CDM}}\)
& \(\boldsymbol{n=2}\)
& \(\boldsymbol{n=3}\)
& \(\boldsymbol{n=4}\)\\
\hline

\multicolumn{5}{l}{\textit{Sampled cosmological parameters}}\\
\(100\omega_b\)
& \margbf{2.263\pm0.013}{2.262}
& \margbf{2.280^{+0.027}_{-0.016}}{2.307}
& \margbf{2.300\pm0.020}{2.306}
& \margbf{2.295\pm0.018}{2.301}\\
\(\omega_{\mathrm{cdm}}\)
& \margbf{0.11710\pm0.00062}{0.11714}
& \margbf{0.1243\pm0.0034}{0.1284}
& \margbf{0.1262\pm0.0026}{0.1275}
& \margbf{0.1249^{+0.0025}_{-0.0013}}{0.1275}\\
\(H_0\)
& \margbf{68.73\pm0.28}{68.69}
& \margbf{70.55^{+0.80}_{-0.66}}{71.68}
& \margbf{71.09\pm0.66}{71.28}
& \margbf{70.81^{+0.69}_{-0.45}}{71.51}\\
\(\tau\)
& \margbf{0.0629^{+0.0071}_{-0.0080}}{0.0610}
& \margbf{0.0604\pm0.0078}{0.0640}
& \margbf{0.0580\pm0.0070}{0.0548}
& \margbf{0.0590\pm0.0068}{0.0575}\\
\(\ln(10^{10}A_{\mathrm{s}})\)
& \margbf{3.056^{+0.014}_{-0.016}}{3.061}
& \margbf{3.066\pm0.017}{3.081}
& \margbf{3.067\pm0.014}{3.062}
& \margbf{3.065\pm0.014}{3.074}\\
\(n_{\mathrm{s}}\)
& \margbf{0.9722\pm0.0033}{0.9727}
& \margbf{0.9809^{+0.0071}_{-0.0052}}{0.9876}
& \margbf{0.9873\pm0.0052}{0.9863}
& \margbf{0.9831\pm0.0045}{0.9881}\\
\multicolumn{5}{l}{\textit{Sampled modified-gravity parameters}}\\
\(\phi_{\mathrm{ini}}\)
& \notapp
& \margbf{0.502^{+0.12}_{-0.064}}{0.604}
& \margbf{0.731^{+0.11}_{-0.089}}{0.783}
& \margbf{>0.839}{0.962}\\
\(\log_{10}V_0\)
& \notapp
& \margbf{3.71^{+0.86}_{-0.59}}{3.81}
& \margbf{3.63^{+0.21}_{-0.31}}{3.52}
& \margbf{3.25^{+0.25}_{-0.34}}{2.96}\\
\(c_1\)
& \notapp
& \notapp
& \notapp
& \margbf{\left(1.7^{+1.1}_{-1.2}\right)\times10^{-4}}{2.9\times10^{-4}}\\
\(c_2\)
& \notapp
& \margbf{>-0.107}{-0.001}
& \margbf{>-0.0695}{-0.0187}
& \margbf{>-0.0178}{-0.0228}\\
\multicolumn{5}{l}{\textit{Derived parameters}}\\
\(\Omega_{\mathrm{m}}\)
& \margbf{0.2972\pm0.0035}{0.2975}
& \margbf{0.2968^{+0.0044}_{-0.0035}}{0.2961}
& \margbf{0.2965\pm0.0034}{0.2976}
& \margbf{0.2962\pm0.0035}{0.2956}\\
\(\Omega_{\Lambda}\)
& \margbf{0.7027\pm0.0035}{0.7024}
& \margbf{0.7032^{+0.0035}_{-0.0044}}{0.7038}
& \margbf{0.7034\pm0.0034}{0.7023}
& \margbf{0.7037\pm0.0035}{0.7044}\\
\(100\omega_{\rm m}\)
& \margbf{14.037\pm0.060}{14.040}
& \margbf{14.77^{+0.46}_{-0.53}}{15.22}
& \margbf{14.98\pm0.27}{15.12}
& \margbf{14.85^{+0.26}_{-0.13}}{15.11}\\
\(\sigma_8\)
& \margbf{0.8071^{+0.0056}_{-0.0065}}{0.8096}
& \margbf{0.830\pm0.011}{0.832}
& \margbf{0.843\pm0.011}{0.837}
& \margbf{0.8323\pm0.0090}{0.8473}\\
\(r_{\mathrm{drag}}\,[\mathrm{Mpc}]\)
& \margbf{147.58\pm0.19}{147.58}
& \margbf{143.7^{+2.5}_{-2.2}}{141.5}
& \margbf{142.7\pm1.3}{142.0}
& \margbf{143.28^{+0.58}_{-1.3}}{141.97}\\
\(\dfrac{G(\phi_{\mathrm{ini}})}{G_0}\)
& \margbf{1}{1}
& \margbf{1.0190^{+0.0079}_{-0.017}}{1.0005}
& \margbf{1.028^{+0.012}_{-0.022}}{1.012}
& \margbf{1.0110^{+0.0031}_{-0.011}}{1.0212}\\
\(S_8\)
& \margbf{0.8033\pm0.0080}{0.8063}
& \margbf{0.826\pm0.013}{0.827}
& \margbf{0.838\pm0.012}{0.834}
& \margbf{0.827\pm0.010}{0.841}\\
\hline\hline
\end{tabular}%
}
\caption{
Marginalized constraints (posterior means and $68\%$ limits) and best-fit values (in parenthesis) for the P-L, P-LB and P-LB-SN datasets. We omit the $c_1$ parameter report in all cases due to finding a bimodal distribution, with the exception of the $n=4$ scenario in the P-LB-SN dataset, which found a single mode. \label{tab:constraints_P_PB}
}
\end{table*}

\subsection{Full data combination P-LB-SN}
\label{subsec:results-pbs}

As expected, the inclusion of Pantheon+ with SH0ES calibration produces significant shifts of the marginalized posteriors when compared to the other joint analyses. Here we summarize the results found for the cosmological parameters, and leave the study of the modified gravity parameters to the next section. 

Table~\ref{tab:H0_rsdrag_S8} presents the constraints on the $H_0$, $r_{\rm drag}$, and $S_8$ parameters. We find a substantial increase in $H_0$ relative to $\Lambda$CDM for the same P-LB-SN data combination, which shift toward the Pantheon+ and SH0ES reference value of $H_0=73.6\pm1.1\ \mathrm{km\,s^{-1}\,Mpc^{-1}}$ \cite{Brout:2022vxf}. The upward shift is accompanied by broader posterior intervals, but the 1$\sigma$ constraints remain at approximately the percent-level or lower, which constitutes an excellent precision. Nevertheless, we note that since the SH0ES calibration is already included in the joint likelihood, this comparison does not constitute an independent measurement of the residual Hubble tension.

\begin{table}[H]
    \centering
    \renewcommand{\arraystretch}{1.3}
    \begin{tabular}{l|c|c|c}
        \hline\hline
        Model & $H_0$ & $r_{\rm drag}$ & $S_8$ \\
        \hline
        $\Lambda$CDM & $68.73\pm 0.28$ & $147.58\pm0.19$ & $0.8033\pm 0.0080$ \\
        $n=2$ & $70.55^{+0.80}_{-0.66}$ & $143.7^{+2.5}_{-2.2}$ & $0.826\pm 0.013$ \\
        $n=3$ & $71.09\pm0.66$ & $142.7\pm1.3$ & $0.838\pm 0.012$ \\
        $n=4$ & $70.81^{+0.69}_{-0.45}$ & $143.28^{+0.58}_{-1.3}$ & $0.827\pm 0.010$ \\
        \hline\hline
    \end{tabular}
    \caption{Summary of results for the $H_0$, $r_{\rm drag}$ and $S_8$ cosmological parameters in the P-LB-SN statistical analysis.}
    \label{tab:H0_rsdrag_S8}
\end{table}

The upward shift in $H_0$ is accompanied by a reduction of the sound horizon, which can be understood from the usual definitions
\begin{equation}
 \begin{aligned}
 r_{\rm drag}=\int_{z_{\rm drag}}^{\infty}
 \frac{c_s(z)}{H(z)}\,\mathrm{d}z, \qquad
 c_s^2=\frac{1}{3(1+R_b)}.
 \end{aligned}
 \label{eq:rdrag-interpretation}
\end{equation}
where $R_b=3\rho_b/4\rho_\gamma$. The P-LB-SN dataset shifts the physical baryon density from approximately $\omega_{\rm b} =0.0226$ in $\Lambda$CDM to $\omega_{\rm b}=0.0230$ in the MG scenarios, therefore increasing the baryon loading and lowering the photon-baryon sound speed. It is not, however, the only contribution: the physical matter density $\omega_{\rm m}$ also rises from approximately $\omega_{\rm m} = 0.140$ to $0.148$--$0.150$, and the unfreezing of the scalar field injects energy into the primordial fluid (see fig.~\ref{fig:bestfit_scalar_background}). All three factors increase the pre-recombination expansion rate and decrease the sound horizon as a consequence.

The full combination also drives the primordial spectrum toward a substantially bluer tilt, with $n_s>0.98$ for all three potential scenarios, while $\ln(10^{10}A_s)$ slightly increases to approximately $3.066$.  This is the familiar compensating response of the high-$\ell$ CMB spectrum when the pre-recombination background is modified \cite{PhysRevD.104.063510}: additional small-scale primordial power helps preserve the observed damping tail.

\begin{figure}[ht]
    \centering
    \includegraphics[width=1\linewidth]{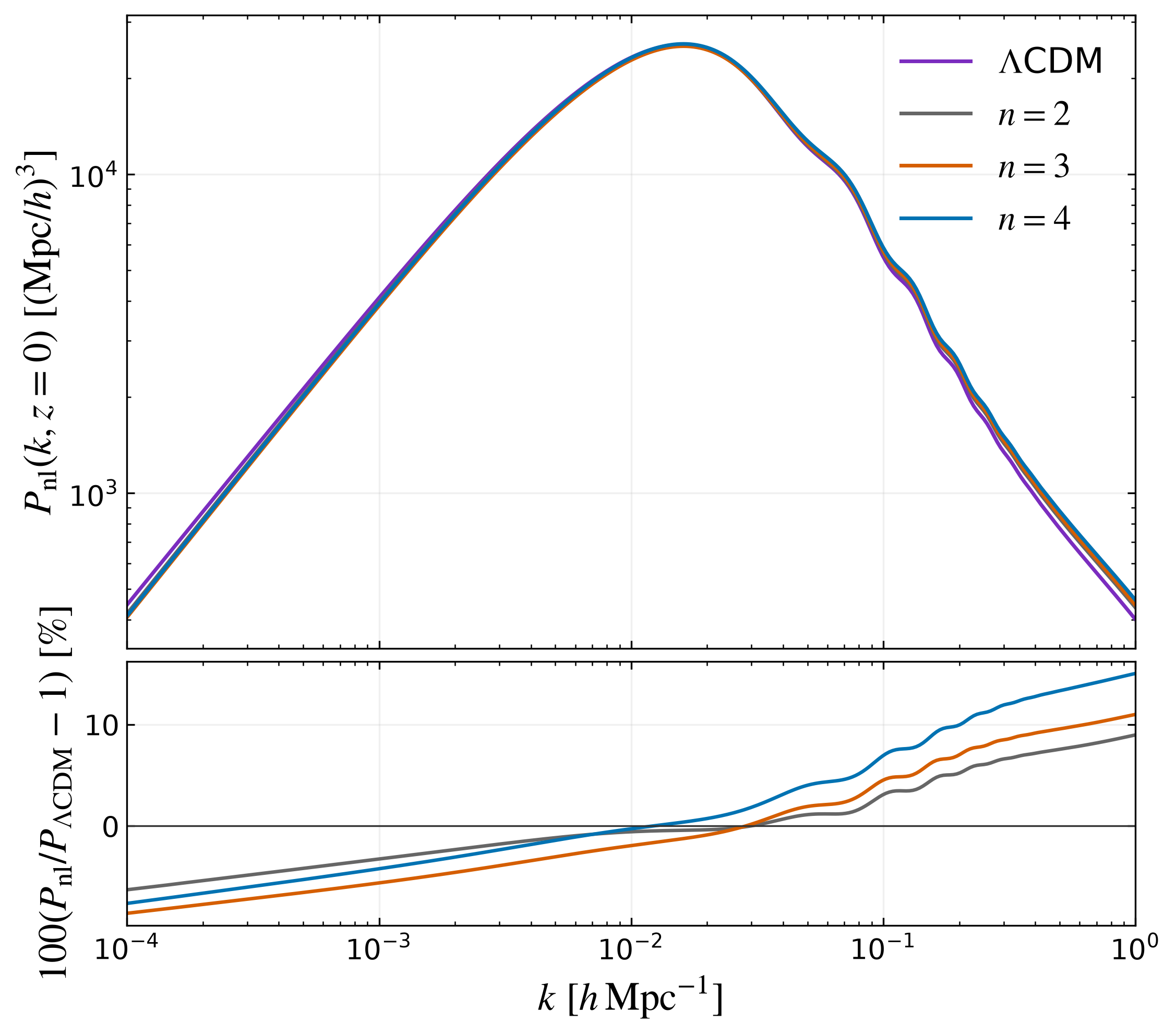}
    \caption{Non-linear matter power spectrum at $z=0$ for the P-LB-SN best-fit points (upper) and their difference from $\Lambda \rm CDM$ (bottom).}
    \label{fig:matter_power}
\end{figure}

These shifts in the matter content and inflationary parameters also affect the matter power spectrum, $P_m(k,z)$, which is shown in fig.~\ref{fig:matter_power} (evaluated at redshift zero) for the best-fit values found for the full P-LB-SN dataset. For the subhorizon scales relevant to matter clustering, the matter power spectrum, with a conventional normalization, can be schematically expressed as \cite{EisensteinHu1998,Ishak2019}
\begin{equation}
    P_m(k,z) \propto A_s \left(\frac{k}{k_*}\right)^{n_s}T^2(k)D^2(z),
\end{equation}
where $T(k)$ is the time-independent transfer function, $D(z)$ is the scale-independent linear growth function, and the dependence on the primordial amplitude $A_s$, spectral index $n_s$ and pivot wavenumber $k_* = 0.05 \ \rm Mpc^{-1}$ are made explicit. This relation is useful for interpreting fig.~\ref{fig:matter_power}. Although the absolute spectra appear nearly identical in the upper logarithmic panel, their ratios reveal scale-dependent differences approaching 15\%. The first important contribution is already present in the primordial spectrum. For the full P-LB-SN best fits, the spectral index rises significantly:
\begin{align*}
    n_s&= 0.9727 && \Lambda{\rm CDM}\\
    n_s
    &=
    0.9876,\;0.9863,\;0.9881,
    \qquad
    &&n=2,3,4,
\end{align*}
respectively. The modified models therefore select a bluer primordial spectrum, explaining the large-scale ordering of the matter spectra in fig.~\ref{fig:matter_power}. In addition to this, the physical dark-matter density increases from $\omega_{\rm cdm}=0.1171$ to $0.1243$--$0.1262$, while the baryon fraction decreases; together with the bluer tilt and slightly larger $A_s$, this raises $\sigma_8$, whereas $\Omega_{\rm m}$ remains essentially unchanged. The larger values of $S_8$ therefore stem from the increase in $\sigma_8$.

\subsection{Modified-gravity parameters and the transition history}
\label{subsec:results-mg}

\begin{figure*}
    \centering
    \includegraphics[width=0.78\textwidth]{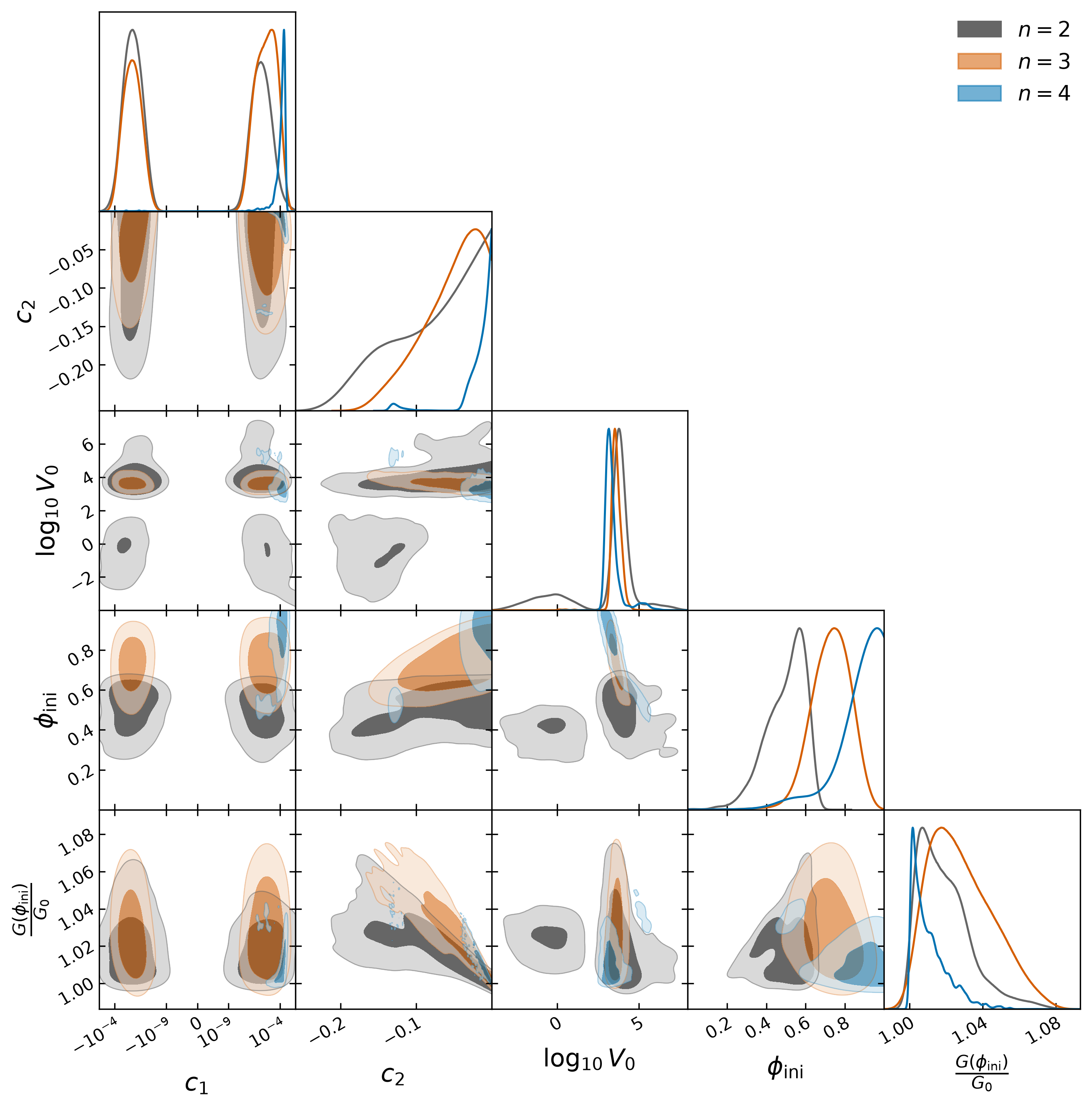}
    \caption{Triangle plot containing the modified-gravity posteriors for P--LB--SN. Contours contain $68\%$ and $95\%$ of the marginalized probability.}
    \label{fig:mg-P-LB-SN}
\end{figure*}

In this section we discuss the statistical constraints on the modified gravity parameters. Given the suppression of the modified sector for the P-L and P-LB datasets, we only present the contour plots for the P-LB-SN case -- see fig.~\ref{fig:mg-P-LB-SN}. It nonetheless qualitatively represents the results from the other likelihood combinations, with the exception of the $\log_{10}V_0$ parameter and the $n=4$ case altogether, which are more restricted relative to P-L and P-LB. 

One feature that clearly stands out is the sharp posterior distribution of the potential amplitude in all models, which is localized at around $\log_{10} V_0 \approx 3.5$. In table~\ref{tab:potential_ev} we show the best-fit full potential and its amplitude in electron-volts units. Noticeably, the potential is concentrated at the $\sim\rm eV^4$ energy scale, which corresponds to the matter-radiation equality epoch. Fig~\ref{fig:bestfit_scalar_background} illustrates this behavior: for the best-fit points, the scalar field unfreezes and injects energy into the cosmological fluid at very similar epochs for all potentials. This follows the standard results from EDE models, which prefer an energy injection occurring just prior to recombination \cite{Poulin_2019, Knox_2020, AGRAWAL2023101347}.

\begin{table}[ht]
    \centering
    \begin{tabular}{l c c c}
    \hline
    \noalign{\vskip 1.5pt}
    \hline

    \multicolumn{1}{c}{ Parameter}
    & \multicolumn{1}{c}{ $n=2$}
    & \multicolumn{1}{c}{ $n=3$}
    & \multicolumn{1}{c}{ $n=4$}
    \\
    \hline

    {$V(\phi_{\rm ini})\,[{\rm eV}^4]$}
    & $0.209$
    & $0.186$
    & $0.163$
    \\

    {$V(\phi_{\rm ini})^{1/4}\,[{\rm eV}]$}
    & $0.676$
    & $0.657$
    & $0.635$
    \\

    \hline
    \end{tabular}
    \caption{Best-fit potentials and their corresponding fourth-root in electron volts units for P--LB--SN.}
    \label{tab:potential_ev}
\end{table}

Moreover, the PolyChord sampler also found a statistically significant second mode for the $n=2$ case, which is more easily observed in the $\log_{10}V_0$ parameter. This mode corresponds to the $1\sigma$ CL in the P-L and P-LB dataset analyses, indicating that it provides a preferable region for the CMB data; enough to make it appear as another mode.

Moreover, we note that $\phi_{\rm ini}$ is pushed to its upper limit in the $n=4$ model. This particular behavior can be understood from the preferred unfreezing epoch. Since
\begin{equation}
    V_{\rm ini}\propto V_0\phi_{\rm ini}^{2n},
    \
    V_{,\phi\phi}(\phi_{\rm ini})\propto  2n(2n-1)V_0\phi_{\rm ini}^{2(n-1)}
    \label{eq:initial-potential-combinations}
\end{equation}
the larger power in the $n=4$ model more strongly suppresses both the initial potential energy and its curvature than the $n=2,3$ cases, as $\phi_{\rm ini}<1$. Therefore, maintaining a sufficiently large pre-recombination contribution with a comparable value of $V_0$ pushes $\phi_{\rm ini}$ toward unity.

This also explains the concentration of the $c_2$ parameter close to zero. Returning to the sub-Planckian field requirement $\phi_{\rm ini}^{2}\leq A(\phi_{\rm ini})<1$. For $\phi_{\rm ini}>0$, this condition implies
\begin{equation}
    \frac{c_1}{\phi_{\rm ini}} - \left(1-\frac{1}{\phi_{\rm ini}^2}\right)\geq |c_2| > \frac{c_1}{\phi_{\rm ini}}.
    \label{eq:c2-induced-bounds}
\end{equation}
As $\phi_{\rm ini}\rightarrow1$, the allowed interval for $A(\phi_{\rm ini})$ collapses toward unity and the two bounds on $c_2$ approach $c_2\approx-|c_1|$. This makes the two coefficients comparable and squeezes $c_2$ toward zero, as the allowed values of $c_1$ are themselves small. For the other models, $\phi_{\rm ini}$ is less constrained and a broader relation between the dimensionless coefficients is permitted.

In regard to the $c_1$ parameter, we note that its posterior contains two approximately mirror branches for $n=2$ and $n=3$, and a localized distribution on the positive branch for $n=4$. The origin of the two $c_1$ branches and the degeneracy breaking of $n=4$ can be qualitatively understood from the dynamical equation for the scalar field given in eq.~\eqref{eq:KG}:
\begin{align*}
    &\ddot{\phi}+3H\dot{\phi} + V_{,\phi} -\frac{1}{2}(c_1+2c_2\phi)R=0.
\end{align*}

Clearly, the sign of $c_1$ affects the dynamic equations, but the prior exploration has shown that $|c_1| \leq 3\times10^{-3}$ (see fig.~\ref{fig:validity_regions}), and so the effects of $c_1$ are suppressed in many cases. In general, the sign of the linear coefficient will be important whenever the modified gravity sector is relevant to the dynamic equations \textit{and} $c_1\phi$ is comparable to (or higher than) $c_2\phi^2$. This is precisely the phenomenon observed in the P-LB-SN results for the $n=4$ model (see fig.~\ref{fig:mg-P-LB-SN}): $c_1$ was driven toward larger values and $c_2$ concentrated near zero, making the two coefficients comparable. But why is the positive branch selected? And why is there a gap between the modes?

To answer these questions, we must first recall that $G_{\rm eff}/G < 1$ necessarily requires $A(\phi)>1$, and since
\begin{equation}
A-1=c_1\phi-|c_2|\phi^2,
\end{equation}
the first weaker-gravity frontier (i.e. the first $\phi_1$ such that $G_{\rm eff}(\phi<\phi_1)/G < 1$) has the same sign as $c_1$. That is, a solution with $c_1>0$ can enter the weaker-gravity interval while the field remains positive, whereas a solution with $c_1<0$ must first cross the origin. Moreover, eq.~\eqref{eq:lower_Geff_bound} shows that, for sufficiently small $c_1$, the depth of $G_{\rm eff}/G$ falls below the machine precision ($\sim 10^{-16}$) used to enforce the weaker-gravity condition, and the corresponding point is rejected. This explains the gap between the two modes of the $c_1$ posterior.

After the scalar field thawing, the restoring force of the $n=4$ potential behaves as $V_{,\phi}\propto V_0\phi^7$, and thus decreases much more rapidly as the field approaches the origin than the corresponding $\phi^3$ and $\phi^5$ forces of the $n=2$ and $n=3$ models. Hence, the Hubble drag becomes more important relative to the rapidly decreasing potential force, which favors a more strongly damped, nearly non-oscillatory approach to $\phi=0$. The field consequently spends less time at the $\phi<0$ branch, or no time at all, strongly suppressing the weaker-gravity histories associated with $c_1<0$. Conversely, the weaker-gravity interval generated by $c_1>0$ lies on the positive-$\phi$ side and can be traversed before the field reaches the origin.

Furthermore, the inferred initial cosmological gravitational strength remains close to GR:
\begin{equation}
 \begin{aligned}
 \frac{G_{\rm N}(\phi_{\rm ini})}{G} &=1.019^{+0.008}_{-0.017} &&(n=2),\\
 &=1.028^{+0.012}_{-0.022} &&(n=3),\\
 &=1.0110^{+0.0031}_{-0.011} &&(n=4).
 \end{aligned}
 \label{eq:Gini-P-LB-SN-results}
\end{equation}
The full-data solutions therefore prefer only low, percent-level early departures from the present gravitational strength.

\begin{figure}[ht]
    \centering
    \includegraphics[width=1\linewidth]{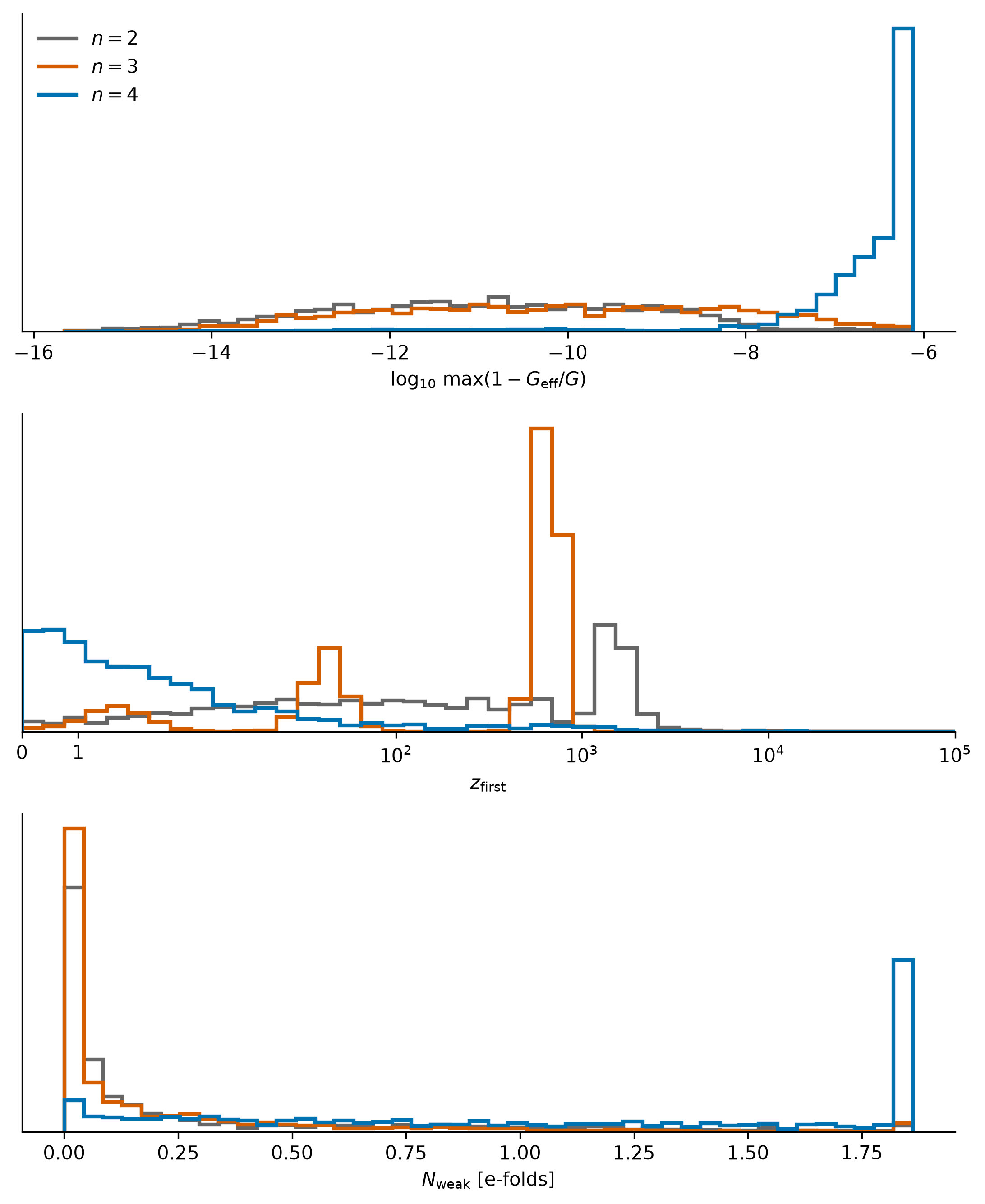}
    \caption{Maximum value of the weaker-gravity excursion, the redshift at which $G_{\rm eff} < G$ first occurs and the total number of e-folds, $N_{\rm weak}$, during which $G_{\rm eff}<G$ for the P-LB-SN dataset.}
    \label{fig:weaker_phase}
\end{figure}

Moreover, our prior choices select histories with $G_{\rm eff}<G_0$ at least once, but impose neither a minimum depth nor a minimum duration for that phase. Fig.~\ref{fig:weaker_phase} shows how the P-LB-SN posteriors make use of this freedom. For $n=2$ and $n=3$, the maximal depth is distributed almost uniformly between $10^{-15}$ and $10^{-7}$, whereas $n=4$ shows a concentration of points on the range $10^{-8}-10^{-6}$. The phase first appears mostly around recombination for $n=2$ ($z_{\rm first}\sim10^3$) and somewhat later for $n=3$ ($z_{\rm first}\sim600$, with a secondary peak at $z\sim40$ and a tertiary peak at $z\sim 1.5$), and is very brief for these two cases, with $N_{\rm weak}\lesssim0.1$. On the other hand, the $n=4$ model is characterized by a very different mechanism. Its maximal depth is concentrated at $1-G_{\rm eff}/G\simeq10^{-8}$--$10^{-6}$, the phase begins late, mostly at $z_{\rm first}\lesssim3$, and lasts up to $N_{\rm weak}\simeq1.9$ e-folds, often until the present. These results are exemplified in fig.~\ref{fig:bestfit_scalar_background}, where $G_{\rm eff}/G$ has sharp spikes with low amplitudes in the $n=2,3$ cases but has a longer excursion on the $n=4$ case.

\subsection{Model Comparison}
\label{subsec:model_comparison}

In this section we compare the statistical performance of the modified gravity models with that of $\Lambda$CDM. We first consider the best-fit effective $\chi^2$, defined by
\begin{equation}
    \chi^2_{\rm eff}=-2\ln\mathcal{L},
    \qquad
    \Delta\chi^2=\chi^2_{\rm MG}-\chi^2_{\Lambda{\rm CDM}},
    \label{eq:model_comparison_chi2}
\end{equation}
where the likelihoods are evaluated at the respective best-fit points for the same data combination, and a negative $\Delta\chi^2$ indicates an improved fit.

Furthermore, to account for the additional complexity of our model due to the inclusion of four additional parameters ($\phi_{\rm ini}$, $\log_{10}V_0$, $c_1^{\rm symlog}$ and $c_2$), we also consider the Bayes factor:
\begin{equation}
    B_{M/\Lambda}=\frac{\mathcal{Z}_M}{\mathcal{Z}_\Lambda},
    \
    \mathcal{Z}_M=\int \mathcal{L}(D|\boldsymbol{\theta},M)
    \pi(\boldsymbol{\theta}|M)\,d\boldsymbol{\theta},
    \label{eq:model_comparison_evidence}
\end{equation}
where $\mathcal{Z}_M$ is the Bayesian evidence and $\pi$ is the normalized parameter prior. The evidence averages the likelihood over the adopted priors and therefore accounts for how much of the available parameter space provides a good fit, making the Bayes factor more informative for model comparison than the best-fit improvement alone~\cite{Trotta_2008}. Thus, we use the global evidences from the processed outputs of the \texttt{Cobaya} code, which retain the normalization of the adopted priors and account for excluded regions\footnote{Cobaya normalizes the internal parameter priors to the full prior box, and its PolyChord interface corrects the processed evidence for regions of null prior or likelihood. See \href{https://cobaya.readthedocs.io/en/latest/sampler_polychord.html\#computing-bayes-ratios}{the Cobaya documentation, ``Computing Bayes ratios''}.}. The results for all models are shown together with the best-fit $\chi^2$ analyses in table~\ref{tab:best_fit_chi2}.

To interpret the Bayes factors, we use the Jeffreys scale adopted by~\cite{Trotta_2008} (summarized in table~\ref{tab:jeffreys_scale}). Positive $\ln B_{M/\Lambda}$ favors modified gravity, while negative values favor $\Lambda$CDM.

For the CMB-only analysis, the $n=3$ and $n=4$ best fits are as good as $\Lambda$CDM ($|\Delta\chi^2|<0.3$), while $n=2$ improves the fit by $\Delta\chi^2=-1.6$. Adding BAO gives $\Delta\chi^2_{\rm tot}=-2.0,-1.7,+0.9$ for $n=2,3,4$, respectively, showing that $\Lambda$CDM already describes these data well and that the additional scalar sector produces little improvement. This justifies the poor Bayes Factor of the scalar-tensor model: for CMB alone, the central values indicate a moderate preference for $\Lambda$CDM over $n=2$ and a weak preference over $n=3,4$, while adding BAO leaves a weak preference for $\Lambda$CDM in all three cases.

On the other hand, the full P-LB-SN combination gives a substantially better fit for all three MG models:
\begin{equation}
    \Delta\chi^2_{\rm tot}=-17.4,\,-18.1,\,-17.6,
    \qquad n=2,3,4,
\end{equation}
respectively. This improvement comes from the supernova likelihood including SH0ES, for which $\Delta\chi^2_{\rm SN}=-22.5,-21.2,-21.1$, and more than compensates for the modest deterioration in the CMB and BAO contributions. Furthermore, all three modified gravity models become favored under the adopted priors. The largest evidence is obtained for $n=3$, with $\ln B_{3/\Lambda}=4.55\pm0.44$ and $B_{3/\Lambda}\simeq94.5$, corresponding to moderate evidence, whereas $n=2,4$ give $\ln B_{M/\Lambda}=2.37\pm0.44,\,2.30\pm0.85$ and Bayes factors near ten, classified as weak evidence by their central values. Thus, the full-data preference survives averaging over the adopted parameter priors.

\begin{table}[ht]
    \centering
    \small
    \setlength{\tabcolsep}{4pt}
    \begin{tabular}{@{}lll@{}}
        \toprule
        $|\ln B|$ & Approximate odds & Interpretation \\
        \midrule
        $<1$            & $<3:1$          & Inconclusive \\
        $[1,2.5)$       & $3:1$--$12:1$   & Weak \\
        $[2.5,5)$       & $12:1$--$150:1$ & Moderate \\
        $\geq5$         & $\geq150:1$     & Strong \\
        \bottomrule
    \end{tabular}
    \caption{Jeffreys scale adopted from~\cite{Trotta_2008}. The odds refer to the evidence ratio in favour of the preferred model.}
    \label{tab:jeffreys_scale}
\end{table}

Our model therefore provides a substantially better fit when the SH0ES calibration is included, but determining whether a statistical preference remains without it requires new parameter fits and evidence calculations for P-LB combined with Pantheon+ alone.

\begin{table*}
    \centering
    \small
    \setlength{\tabcolsep}{4pt}
    \renewcommand{\arraystretch}{1.2}
    \begin{tabular}{llcccccc}
        \toprule
        Data & Model
        & \shortstack{$\chi^2_{\mathrm{CMB}}$\\$(\Delta\chi^2_{\mathrm{CMB}})$}
        & \shortstack{$\chi^2_{\mathrm{BAO}}$\\$(\Delta\chi^2_{\mathrm{BAO}})$}
        & \shortstack{$\chi^2_{\mathrm{SN}}$\\$(\Delta\chi^2_{\mathrm{SN}})$}
        & \shortstack{$\chi^2_{\mathrm{tot}}$\\$(\Delta\chi^2_{\mathrm{tot}})$}
        & $\ln B_{M/\Lambda}$ & $B_{M/\Lambda}$ \\
        \midrule
        P-L & $\Lambda$CDM
        & $1011.6\;(0.0)$ & -- & --
        & $1011.6\;(0.0)$ & $0$ & $1$ \\
         & $n=2$
        & $1010.0\;(-1.6)$ & -- & --
        & $1010.0\;(-1.6)$ & $-3.03\pm0.42$ & $0.048$ \\
         & $n=3$
        & $1011.5\;(0.0)$ & -- & --
        & $1011.5\;(0.0)$ & $-2.00\pm0.42$ & $0.135$ \\
         & $n=4$
        & $1011.3\;(-0.2)$ & -- & --
        & $1011.3\;(-0.2)$ & $-2.30\pm0.42$ & $0.100$ \\
        \midrule
        P-LB & $\Lambda$CDM
        & $1014.0\;(0.0)$ & $11.8\;(0.0)$ & --
        & $1025.8\;(0.0)$ & $0$ & $1$ \\
         & $n=2$
        & $1012.5\;(-1.4)$ & $11.2\;(-0.6)$ & --
        & $1023.7\;(-2.0)$ & $-1.26\pm0.43$ & $0.283$ \\
         & $n=3$
        & $1013.6\;(-0.4)$ & $10.4\;(-1.3)$ & --
        & $1024.0\;(-1.7)$ & $-1.30\pm0.43$ & $0.273$ \\
         & $n=4$
        & $1015.3\;(+1.3)$ & $11.3\;(-0.5)$ & --
        & $1026.6\;(+0.9)$ & $-2.00\pm0.43$ & $0.135$ \\
        \midrule
        P-LB-SN & $\Lambda$CDM
        & $1015.8\;(0.0)$ & $10.5\;(0.0)$ & $1483.2\;(0.0)$
        & $2509.5\;(0.0)$ & $0$ & $1$ \\
         & $n=2$
        & $1020.4\;(+4.6)$ & $11.0\;(+0.5)$ & $1460.7\;(-22.5)$
        & $2492.1\;(-17.4)$ & $+2.37\pm0.44$ & $10.7$ \\
         & $n=3$
        & $1018.1\;(+2.3)$ & $11.2\;(+0.7)$ & $1462.0\;(-21.2)$
        & $2491.4\;(-18.1)$ & $+4.55\pm0.44$ & $94.5$ \\
         & $n=4$
        & $1019.2\;(+3.4)$ & $10.7\;(+0.1)$ & $1462.1\;(-21.1)$
        & $2492.0\;(-17.6)$ & $+2.30\pm0.85$ & $10.0$ \\
        \bottomrule
    \end{tabular}
    \caption{Best-fit effective $\chi^2$ contributions and Bayesian evidence
    comparisons for the different models and data combinations.     Parenthetical values give $\Delta\chi^2_X\equiv\chi^2_X-\chi^2_{X,\Lambda\mathrm{CDM}}$, relative to the $\Lambda$CDM best fit to the same data combination, computed before rounding.}
    \label{tab:best_fit_chi2}
\end{table*}

\section{Conclusion}

We have investigated a generalized Brans--Dicke scalar-tensor model with a polynomial non-minimal coupling $A(\phi)=1+c_1\phi+c_2\phi^2$ and power-law potentials $V(\phi)\propto V_0\phi^{2n}$, for $n=2,3,4$.

In particular, we study the parameter space for which $G_\text{N} > G$ before and around recombination and $G_{\rm eff} < G$ at least once during the subsequent evolution; we do not impose a minimum amplitude, duration or epoch of the weaker gravity regime. This gravity-strength transition allows the expansion rate to be greater in the early-Universe (hence decreasing the sound horizon) and the matter clustering to weaken at late-times when compared to the $\Lambda$CDM model, which could simultaneously increase $H_0$ while decreasing $S_8$ relative to it.

To further limit the theory to more desirable behaviors, we applied the existing local and BBN observational bounds and also imposed that the field is sub-Planckian and that its effective fluid respects the Dominant Energy Condition and has no ghosts. Under these conditions, we find that the linear coupling coefficient is restricted to $|c_1| \lesssim 3\times 10^{-3}$, while the quadratic coupling is strictly negative $c_2 < 0$ once a weaker-gravity regime is required. Nevertheless, the potential, $V_0$, still spans several orders of magnitude (probed instead with $\log_{10}V_0$) and the initial value of the scalar field, $\phi_{\rm ini}$, covers the full sub-Planckian range. Hence, the prior conditions on the theory do not greatly constrain these two parameters.

Moreover, the early stronger-gravity and late weaker-gravity phases compete for the same coefficients, so that a percent-level early enhancement of the gravitational coupling limits the weaker-gravity regime to $1-G_{\rm eff}/G\lesssim10^{-4}$, in agreement with the maximum depth of $2\times10^{-4}$ found in the Sobol exploration. However, no limitation on the duration or thawing epoch for the weaker-gravity regime could be found by local observations alone, and so we confronted the theory with cosmological data from Planck 2018 temperature, polarization and lensing, DESI DR2 BAO and Pantheon$+$ supernova with SH0ES calibration.

In summary, we find that the Planck 2018 and DESI DR2 BAO data cannot properly distinguish between the MG model and $\Lambda$CDM, and so all MG parameters remain relatively unconstrained with respect to the priors. With CMB data alone, the scalar sector is suppressed either by large potential amplitudes, which make the field thaw and decay well before recombination ($n=3,4$), or by small initial field values ($n=2$); adding DESI DR2 BAO favors lower potentials and allows extended high-$H_0$ tails for $n=2,3$. On the other hand, including Pantheon+ with SH0ES calibration raises the inferred $H_0$ and lowers the sound horizon substantially, with the best-fit scalar contributions peaking around matter--radiation equality. For example, for $n=3$, we obtain $H_0=71.09\pm0.66\,\mathrm{km\,s^{-1}\,Mpc^{-1}}$ and $r_{\rm drag}=142.7\pm1.3\,\mathrm{Mpc}$, compared with $H_0=68.73\pm0.28\,\mathrm{km\,s^{-1}\,Mpc^{-1}}$ in $\Lambda$CDM for the same data. The full combination of datasets improves the best-fit effective $\chi^2$ relative to $\Lambda$CDM by $\Delta\chi^2_{\rm tot}=-16.4,-17.5,-17.0$ for $n=2,3,4$, respectively, which occurs predominantly due to the supernova likelihood including SH0ES calibration; this compensates for the small deterioration in the CMB and BAO contributions. The computed Bayesian evidence is largest for $n=3$, with $\ln B_{3/\Lambda}=4.55\pm0.44$, corresponding to a moderate preference on the adopted scale, whereas P-L and P-LB favor $\Lambda$CDM. The statistical preference for the model is therefore tied to the SH0ES calibration.

Furthermore, the full data combination realizes the weaker-gravity regime in two distinct ways: brief excursions after recombination for $n=2,3$, lasting $N_{\rm weak}\lesssim0.1$ e-folds, and, for $n=4$, as an abrupt late-time transition, mostly at $z\lesssim3$, after which $G_{\rm eff}$ remains slightly below $G$, often until the present. In all cases its depth stays at $1-G_{\rm eff}/G\lesssim10^{-6}$, far too shallow to affect the growth of structure. In fact, all three potentials give higher $S_8$ than $\Lambda$CDM when all data sets are considered, which is caused by higher values of $\omega_{\rm cdm}$, $n_s$ and, to a lesser extent, $A_s$. The possibility of a gravitational transition therefore does not translate into a reduction of $S_8$ within the present analysis, and the preferred cosmologies are of the early-modified-gravity type.

Therefore, the proposed generalized Brans--Dicke model is theoretically consistent and competitive with $\Lambda$CDM when the local distance ladder is included, even with 4 extra parameters, but within the conditions adopted here its weaker-gravity regime cannot be deep enough to affect the growth of structure. Further studies could clarify these conclusions: an analysis with Pantheon+ without the SH0ES calibration, to determine whether the preference survives without the distance ladder; the inclusion of the ACT DR6 \cite{louis2025atacamacosmologytelescopedr6} and SPT-3G high-$\ell$ CMB data \cite{Camphuis:2025spt}, together with data on Large Scale Structure; and the relaxation of some of the theoretical conditions, such as the energy-condition prior or the quadratic truncation of the coupling function, which could allow the two gravitational phases to be controlled independently.

\begin{acknowledgments}
LPCL thanks the Coordena\c{c}\~ao de Aperfei\c{c}oamento de Pessoal de N\'ivel Superior -- Brasil (CAPES) -- Finance Code 001, for the partial funding of this reasearch. GCD would like to thank CNPq for financial support under grant no. 307565/2025-4. This research was funded by the Spanish Grant PID2023-149560NB-C21 and the Severo Ochoa Excellence Grant CEX2023-001292-S, funded by MICIU/AEI/10.13039/501100011033 (``ERDF A way of making Europe'', ``PGC Generacion de Conocimiento'') and FEDER, UE. Support from CosmoVerse CA21136 and CaLISTA CA21109 COST actions, European Cooperation in Science and Technology is also acknowledged. This work has also been supported by the European Horizon Europe staff exchange (SE) programme HORIZON-MSCA2021-SE-01 Grant No. NewFunFiCO-101086251.
\end{acknowledgments}

\section*{Data Availability}
The \texttt{Cobaya} input files and the resulting nested-sampling chains supporting the findings of this study are available from the corresponding author upon reasonable request.

\appendix

\section{hi\_class configuration}\label{app:hi_class}

The \texttt{hi\_class} configuration settings are presented in Listing \ref{lst:class_params}.

\begin{lstlisting}[caption={CLASS configuration and high-precision parameters.}, label={lst:class_params}]
N_ur: 2.0328
N_ncdm: 1
m_ncdm: 0.06
T_ncdm: 0.71611
T_cmb: 2.7255
P_k_max_h/Mpc: 1.0
non_linear: halofit
output: lCl tCl pCl mPk
modes: s
lensing: 'yes'
gravity_model: st_theory
Omega_fld: 0.0
Omega_smg_debug: 1.0e-10
output_background_smg: 10
skip_stability_tests_smg: 'yes'
pert_initial_conditions_smg: gravitating_attr
tol_background_integration: 5e-4
pert_ic_tolerance_smg: 0.0
use_pert_var_deltaphi_smg: y
recombination: HyRec
accurate_lensing: 1
perturbations_sampling_stepsize: 0.05
l_logstep: 1.025
l_linstep: 20
delta_l_max: 1800
l_switch_limber: 30.0
hyper_sampling_flat: 32.0
l_max_g: 40
l_max_ur: 35
l_max_pol_g: 60
ur_fluid_approximation: 2
ur_fluid_trigger_tau_over_tau_k: 130.0
radiation_streaming_approximation: 2
radiation_streaming_trigger_tau_over_tau_k: 240.0
hyper_flat_approximation_nu: 7000.0
transfer_neglect_delta_k_S_t0: 0.17
transfer_neglect_delta_k_S_t1: 0.05
transfer_neglect_delta_k_S_t2: 0.17
transfer_neglect_delta_k_S_e: 0.17
start_small_k_at_tau_c_over_tau_h: 0.0001
start_large_k_at_tau_h_over_tau_k: 0.0001
tight_coupling_trigger_tau_c_over_tau_h: 0.005
tight_coupling_trigger_tau_c_over_tau_k: 0.008
start_sources_at_tau_c_over_tau_h: 0.006
l_max_ncdm: 30
tol_ncdm_synchronous: 1.0e-06
\end{lstlisting}

We note that the highly oscillatory nature of the scalar field demands strict precision settings for the background integration. The precision parameters are common to all potential cases. We found that in many regions, the fixed tolerance induced integration step sizes that were too small and caused the differential equation solver to stall or crash entirely, and so such region could be explored by relaxing the tolerance value. On the other hand, relaxing the tolerance leads to a lower numerical reliability, whose effect on the results must be carefully accounted for. In principle, these tolerances could be made dynamical (i.e. dependent on the modified gravity parameters proposed) to further extend the parameter space, but here we set a fixed value that compromises between numerical stability and the volume of the parameter space that can be reliably explored.

\bibliographystyle{JHEP}
\bibliography{references}

\end{document}